\documentclass[iop]{emulateapj}
\usepackage[citebordercolor={0 .5 .5}]{hyperref}

\newcommand{\mum}{\ensuremath{\mu \mathrm{m}}}
\newcommand{\fpsw}{\ensuremath{S_{250\mum}}}
\newcommand{\fpmw}{\ensuremath{S_{350\mum}}}
\newcommand{\fplw}{\ensuremath{S_{500\mum}}}

\begin{document}

\title{HerMES: Candidate
high-redshift galaxies discovered with {\it Herschel}/SPIRE$^{\dagger}$}
\shorttitle{SPIRE high-z  candidates}
\shortauthors{Dowell et al.}

\author{C.~Darren~Dowell\altaffilmark{1,2},
A.~Conley\altaffilmark{3},
J.~Glenn\altaffilmark{4,3},
V.~Arumugam\altaffilmark{5},
V.~Asboth\altaffilmark{6},
H.~Aussel\altaffilmark{7},
F.~Bertoldi\altaffilmark{8},
M.~B{\'e}thermin\altaffilmark{7},
J.~Bock\altaffilmark{1,2},
A.~Boselli\altaffilmark{9},
C.~Bridge\altaffilmark{1},
V.~Buat\altaffilmark{9},
D.~Burgarella\altaffilmark{9},
A.~Cabrera-Lavers\altaffilmark{10,11,12},
C.M.~Casey\altaffilmark{13},
S.C.~Chapman\altaffilmark{14,15},
D.L.~Clements\altaffilmark{16},
L.~Conversi\altaffilmark{17},
A.~Cooray\altaffilmark{18,1},
H.~Dannerbauer\altaffilmark{19},
F.~De Bernardis\altaffilmark{18},
T.P.~Ellsworth-Bowers\altaffilmark{4},
D.~Farrah\altaffilmark{20},
A.~Franceschini\altaffilmark{21},
M.~Griffin\altaffilmark{22},
M.A.~Gurwell\altaffilmark{23},
M.~Halpern\altaffilmark{6},
E.~Hatziminaoglou\altaffilmark{24},
S.~Heinis\altaffilmark{9},
E.~Ibar\altaffilmark{25,26},
R.J.~Ivison\altaffilmark{25,5},
N.~Laporte\altaffilmark{10,11},
L.~Marchetti\altaffilmark{27,21},
P.~Mart{\'i}nez-Navajas\altaffilmark{10,11},
G.~Marsden\altaffilmark{6},
G.E.~Morrison\altaffilmark{13,28},
H.T.~Nguyen\altaffilmark{2,1},
B.~O'Halloran\altaffilmark{16},
S.J.~Oliver\altaffilmark{29},
A.~Omont\altaffilmark{30},
M.J.~Page\altaffilmark{31},
A.~Papageorgiou\altaffilmark{22},
C.P.~Pearson\altaffilmark{32,27},
G.~Petitpas\altaffilmark{23},
I.~P{\'e}rez-Fournon\altaffilmark{10,11},
M.~Pohlen\altaffilmark{22},
D.~Riechers\altaffilmark{33,1},
D.~Rigopoulou\altaffilmark{32,34},
I.G.~Roseboom\altaffilmark{29,5},
M.~Rowan-Robinson\altaffilmark{16},
J.~Sayers\altaffilmark{1},
B.~Schulz\altaffilmark{1,35},
Douglas~Scott\altaffilmark{6},
N.~Seymour\altaffilmark{36},
D.L.~Shupe\altaffilmark{1,35},
A.J.~Smith\altaffilmark{29},
A.~Streblyanska\altaffilmark{10,11},
M.~Symeonidis\altaffilmark{31},
M.~Vaccari\altaffilmark{21,37},
I.~Valtchanov\altaffilmark{17},
J.D.~Vieira\altaffilmark{1},
M.~Viero\altaffilmark{1},
L.~Wang\altaffilmark{29},
J.~Wardlow\altaffilmark{18},
C.K.~Xu\altaffilmark{1,35},
M.~Zemcov\altaffilmark{1}}

\altaffiltext{$\dagger$}{{\it Herschel} is an ESA space observatory with
  science instruments provided by European-led Principal Investigator
  consortia and with important participation from NASA.}
\altaffiltext{1}{California Institute of Technology, 1200 E. California Blvd., Pasadena, CA 91125; cdd@phobos.caltech.edu}
\altaffiltext{2}{Jet Propulsion Laboratory, 4800 Oak Grove Drive, Pasadena, CA 91109}
\altaffiltext{3}{Center for Astrophysics and Space Astronomy 389-UCB, University of Colorado, Boulder, CO 80309}
\altaffiltext{4}{Dept. of Astrophysical and Planetary Sciences, CASA 389-UCB, University of Colorado, Boulder, CO 80309}
\altaffiltext{5}{Institute for Astronomy, University of Edinburgh, Royal Observatory, Blackford Hill, Edinburgh EH9 3HJ, UK}
\altaffiltext{6}{Department of Physics \& Astronomy, University of British Columbia, 6224 Agricultural Road, Vancouver, BC V6T~1Z1, Canada}
\altaffiltext{7}{Laboratoire AIM-Paris-Saclay, CEA/DSM/Irfu - CNRS - Universit\'e Paris Diderot, CE-Saclay, pt courrier 131, F-91191 Gif-sur-Yvette, France}
\altaffiltext{8}{Argelander Institute f\"ur Astronomy, Universit\"at Bonn, Auf dem H\"ugel 71, 53121 Bonn, Germany}
\altaffiltext{9}{Aix-Marseille Universit\'e, CNRS, LAM (Laboratoire d'Astrophysique de Marseille) UMR7326, 13388, France}
\altaffiltext{10}{Instituto de Astrof{\'\i}sica de Canarias (IAC), E-38200 La Laguna, Tenerife, Spain}
\altaffiltext{11}{Departamento de Astrof{\'\i}sica, Universidad de La Laguna (ULL), E-38205 La Laguna, Tenerife, Spain}
\altaffiltext{12}{GTC Project, E-38205 La Laguna, Tenerife, Spain}
\altaffiltext{13}{Institute for Astronomy, University of Hawaii, 2680 Woodlawn Drive, Honolulu, HI 96822}
\altaffiltext{14}{Department of Physics and Atmospheric Science, Dalhousie University, Halifax, NS B3H 3J5, Canada}
\altaffiltext{15}{Institute of Astronomy, University of Cambridge, Madingley Road, Cambridge CB3 0HA, UK}
\altaffiltext{16}{Astrophysics Group, Imperial College London, Blackett Laboratory, Prince Consort Road, London SW7 2AZ, UK}
\altaffiltext{17}{Herschel Science Centre, European Space Astronomy Centre, Villanueva de la Ca\~nada, 28691 Madrid, Spain}
\altaffiltext{18}{Dept. of Physics \& Astronomy, University of California, Irvine, CA 92697}
\altaffiltext{19}{Universit\"at Wien, Institut f\"ur Astrophysic, T\"urkenschanzstra\ss e 17, 1180, Wien, Austria}
\altaffiltext{20}{Department of Physics, Virginia Tech, Blacksburg, VA 24061}
\altaffiltext{21}{Dipartimento di Fisica e Astronomia, Universit\`{a} di Padova, vicolo Osservatorio, 3, 35122 Padova, Italy}
\altaffiltext{22}{School of Physics and Astronomy, Cardiff University, Queens Buildings, The Parade, Cardiff CF24 3AA, UK}
\altaffiltext{23}{Harvard-Smithsonian Center for Astrophysics, 60 Garden Street, Cambridge, MA 02138}
\altaffiltext{24}{ESO, Karl-Schwarzschild-Str. 2, 85748 Garching bei M\"unchen, Germany}
\altaffiltext{25}{UK Astronomy Technology Centre, Royal Observatory, Blackford Hill, Edinburgh EH9 3HJ, UK}
\altaffiltext{26}{Universidad Cat\'olica de Chile, Departamento de Astronom\'ia y Astrof\'isica, Vicu\~na Mackenna 4860, Casilla 306, Santiago 22, Chile}
\altaffiltext{27}{Department of Physical Sciences, The Open University, Milton Keynes MK7 6AA, UK}
\altaffiltext{28}{Canada-France-Hawaii Telescope, Kamuela, HI, 96743}
\altaffiltext{29}{Astronomy Centre, Dept. of Physics \& Astronomy, University of Sussex, Brighton BN1 9QH, UK}
\altaffiltext{30}{Institut d'Astrophysique de Paris, UMR 7095, CNRS, UPMC Univ. Paris 06, 98bis boulevard Arago, F-75014 Paris, France}
\altaffiltext{31}{Mullard Space Science Laboratory, University College London, Holmbury St. Mary, Dorking, Surrey RH5 6NT, UK}
\altaffiltext{32}{RAL Space, Rutherford Appleton Laboratory, Chilton, Didcot, Oxfordshire OX11 0QX, UK}
\altaffiltext{33}{Department of Astronomy, Space Science Building, Cornell University, Ithaca, NY, 14853-6801}
\altaffiltext{34}{Department of Astrophysics, Denys Wilkinson Building, University of Oxford, Keble Road, Oxford OX1 3RH, UK}
\altaffiltext{35}{Infrared Processing and Analysis Center, MS 100-22, California Institute of Technology, JPL, Pasadena, CA 91125}
\altaffiltext{36}{CSIRO Astronomy \& Space Science, PO Box 76, Epping, NSW 1710, Australia}
\altaffiltext{37}{Astrophysics Group, Physics Department, University of the Western Cape, Private Bag X17, 7535, Bellville, Cape Town, South Africa}

\begin{abstract}
We present a method for selecting $z>4$ dusty, star forming galaxies
(DSFGs) using {\it Herschel}/SPIRE 250/350/500\,\mum\ flux densities
to search for red sources.  We apply this method to 21 deg$^2$ of data
from the HerMES survey to produce a catalog of 38 high-$z$ candidates.
Follow-up of the first 5 of these sources confirms that this method is
efficient at selecting high-$z$ DSFGs, with 4/5 at $z=4.3$ to $6.3$
(and the remaining source at $z=3.4$), and that they are some of the
most luminous dusty sources known.  Comparison with previous
DSFG samples, mostly selected at longer wavelengths (e.g.,
  850\,\mum) and in single-band surveys, shows that our method is much
more efficient at selecting high-$z$ DSFGs, in the sense that a much
larger fraction are at $z>3$.  Correcting for the selection
completeness and purity, we find that the number of bright ($\fplw \ge
30$\,mJy), red {\it Herschel} sources is $3.3 \pm 0.8$ deg$^{-2}$.
This is much higher than the number predicted by current models,
suggesting that the DSFG population extends to higher redshifts than
previously believed.  If the shape of the luminosity function for
high-$z$ DSFGs is similar to that at $z\sim2$, rest-frame UV based
studies may be missing a significant component of the star formation
density at $z=4$ to $6$, even after correction for extinction.
\end{abstract}



\section{Introduction}
\nobreak 
The study of massive, dusty, star-forming galaxies (DSFGs) since their
discovery more than a decade ago \citep{smail97, hughes98, barger98} has
fundamentally changed our understanding of the cosmic history of star
formation and galaxy evolution \citep[e.g.,][]{lagache05}.  These
sources are generally believed to be the progenitors of massive
elliptical galaxies in the current epoch.  They were first studied in
the sub-mm and mm\footnote{Classic sub-millimeter galaxies (SMGs) are
  simply DSFGs selected at $\sim 850\,\mum$.}, where they have the
remarkable property that their observed brightness at a fixed
luminosity is almost independent of redshift over roughly $1<z<10$
\citep{blain93, blain02} due to the shape of their spectral energy
distributions (SEDs), an effect which is known as a negative
$K$-correction.  The technology and instrumentation to exploit this
advantage is challenging, however, and ground-based sub-mm/mm
instruments have typically only been able to map areas down to depths
of a few mJy over hundreds of arcminutes$^2$ through narrow
atmospheric windows \citep{Eales:00, scott02, borys03, greve04,laurent05,
  coppin06, bertoldi07, per08, scott08, weiss09, austermann10,
Aretxaga:2011}.

Obtaining redshifts for these objects is a painstaking process
\citep[see e.g.,][]{chapman05}. The most common technique up to this
point, which relies on identifying the radio or mid-IR counterparts
\citep[e.g.,][]{ivison07, ros10} to provide sufficiently precise localizations
and the source being sufficiently bright for successful optical
spectroscopy, works relatively poorly at $z>3$ because such
observations do not benefit from negative $K$-corrections.
Furthermore, even for very high-$z$ sources, ground-based sub-mm/mm
observations are generally limited to probing only the red,
Rayleigh-Jeans side of the thermal SED, and hence can only provide
extremely crude redshift estimates.

As a result, until recently the number of known $z>4$ DSFGs was
relatively limited, although there were some (mostly photometric)
hints that the redshift distribution extended beyond $z=3$
\citep[e.g.][]{dannerbauer:02, younger:07, younger:09}. However, the
known $z>4$ DSFGs were selected in a fairly irregular fashion, making
it difficult to place any quantitative limits on the number of such
sources.  Theoretically, the existence of even the lower-$z$ DSFGs ($z
\approx 2.5$) has proven somewhat challenging to explain
\citep[e.g.,][]{baugh05}. The primary challenge is to accrete enough
gas into the center of massive dark matter halos at early times in
order to fuel these starbursts.  These difficulties are only
exacerbated at higher redshifts because the number of massive galaxies
is expected to decrease rapidly at high-$z$ \citep{hayward:13}.
Therefore, a significant population of $z>4$ massive starbursts would
be a significant challenge to models.  Despite this, there are some
indirect lines of evidence that suggest that the most massive galaxies
may have formed stars at such early times.  In contrast to the bulk of
the field galaxy population, for which most star formation occurs
after $z=2$ \citep[e.g.,][]{sobral12}, studies of the $K$-band
luminosity function of clusters show that the stellar mass in the
brightest members is already in place earlier \citep{capozzi12}.
Detailed study of individual sources favors a star formation epoch of
$z > 4$ \citep{stott11, kaviraj13}.  However, what is missing is an
effective technique for selecting such sources directly while they are
experiencing a starburst.

The {\it Herschel Space Observatory} \citep{pilbratt10}, which
observed at multiple bands spanning the peak of the SED at effectively
all redshifts, and mapped much larger areas down to the confusion
limit than previous surveys, could measure $T_{\mathrm{dust}}/(1+z)$
for a large number of individually detected sources. Since most known
distant DSFGs have dust temperatures in the range $\sim20$--$80$\,K
\citep[e.g.,][]{kovacs06, Casey:12}, the observed {\it Herschel}
colors can be used to select potential high-$z$ sources.  In this
paper we use a map-based technique to search for red, and hence
potentially $z>4$, sources in 21.4 deg$^2$ within the Science
Demonstration Phase fields from the
HerMES\footnote{\url{http://hermes.sussex.ac.uk}} project
\citep{hermes12} at 250, 350, and 500\,\mum . In particular, we select
``red'' sources with $\fplw \geq \fpmw \geq \fpsw$ -- so-called
500\,\mum-risers -- to provide a catalog of high-$z$ DSFGs candidates
with $\fplw \geq 30$ mJy.  The highlight of this initial catalog was
the discovery of the highest-$z$ massive starburst to date, a $z=6.34$
source with a star formation rate of $3000\,\mathrm{M}_{\odot}
\mathrm{yr}^{-1}$ \citep[FLS 3;][]{rie13}.

We present follow-up of some of these sources, including redshifts for
four additional targets from our sample. We find that four of five
500\,\mum-risers selected with our technique and with measured
redshifts lie at $z>4$ (and the remaining source is at $z=3.4$),
demonstrating that this technique is an effective method of selecting
high-$z$ DSFGs.  Returning to the full catalog, we characterize our
selection using completeness and purity simulations in order to
measure the number density of red sources (\S\ref{sec:numdens}), show
that existing literature models in general significantly under-predict
the number of bright, red sources we find
(\S\ref{subsec:modelcompare}), and estimate their contribution to the
star formation history of the Universe (\S\ref{subsec:contribsfr}).
Comparison with other surveys of our fields shows that our sources are
redder than typical sources selected at 1\,mm
(\S\ref{subsec:mmoverlap}).

\section{SPIRE Observations and Data Analysis}
\nobreak 
In this paper, we analyze maps of three extragalactic fields observed
as part of the HerMES program with the Spectral and Photometric
Imaging Receiver \citep[][SPIRE]{gri10} on board {\it Herschel}: {\it
  Spitzer} FLS, GOODS North, and the Lockman Hole.  The latter has
been further subdivided based on mapped depth into the North, East,
and SWIRE regions (Table~\ref{tbl:fields}).  SPIRE observed
simultaneously at 250, 350, and 500\,\mum . Although the fields were
selected based on availability of {\it Herschel} Science Demonstration
Phase (SDP) data, all available data for these fields, including those
beyond the SDP, have been used in the generation of our maps.  The
basic observation and calibration procedures are described by
\citet{gri10, Bendo:13}.

Under the hypothesis that redshift dominates temperature evolution in
producing the observed SEDs of distant galaxies, we select objects
which have ``red'' SPIRE colors in the hope of forming a high-redshift
sample.  This should not be expected to select {\it all} such sources.
For example, the $z=3$ lensed DSFG LSW~1 \citep{Conley:11} would not
be red in the SPIRE bands unless it were at $z>7$.

\begin{table*}[t!]
\centering
\caption{Summary of fields used in this analysis}
\begin{tabular}{l|llllll}
\hline
 Field & RA & Dec & Area & $\sigma_{250\,\mum}$ &
 $\sigma_{350\,\mum}$ & $\sigma_{500\,\mum}$ \\
       & (deg) & (deg) & (deg$^2$) & (mJy) & (mJy) & (mJy) \\
\hline\hline
GOODS-N       & 189.23 & 62.22 & 0.33 & 0.8 & 0.8 & 1.0 \\
Lockman-East  & 163.14 & 57.49 & 0.34 & 3.6 & 3.6 & 4.0 \\
Lockman-North & 161.49 & 59.01 & 0.46 & 3.6 & 3.5 & 4.1 \\
FLS           & 259.05 & 59.29 & 6.83 & 5.1 & 5.3 & 6.5 \\
Lockman-SWIRE & 161.68 & 57.97 & 13.46 & 4.2 & 4.1 & 5.3 \\
\hline
\end{tabular}
\tablecomments{
Properties of the fields used in this analysis.  The RA and Dec are
the approximate field centers (J2000), the area excludes regions
masked due to the lack of cross-scans.  The depths are the approximate
1$\sigma$ instrument noise in mJy/beam in the center of each field,
not including confusion noise.  The Lockman-East and Lockman-North
regions are excluded from Lockman-SWIRE. The rms due to confusion in
the SPIRE bands is $\sim 6$ mJy/beam \citep{ngu10}.
\label{tbl:fields}}
\end{table*}

A natural question is why we do not simply search for targets using
pre-existing single-band HerMES catalogs by imposing our color and
minimum flux density requirements. Catalogs have been produced for
these fields in the SPIRE bands which provide a suitable list of
500\,\mum\ sources \citep{smi11,ros10}.  However, at present such
catalogs are not optimal for our selection purposes.  In
many cases, they require detection in the other SPIRE bands or at an
alternate wavelength (e.g., 24\,\mum ) -- requirements which will bias
against red and therefore, for our purposes, interesting sources.
Furthermore, even in those fields where 500\,\mum\ selected catalogs
are available, detailed inspection and tests on simulated data show
that catalog selection is currently both significantly less pure
(i.e., a larger portion of the detected sources are not, in fact, red)
and less complete than the map-based method described below.

\subsection{Map Construction}
\label{subsec:map-construction}
\nobreak 
Our map-based search uses products from the SPIRE-HerMES Iterative Map
maker \citep[SHIM;][]{lev10, viero12}. Briefly, we use redundant
sampling of the same position on the sky with multiple scan directions
(cross-scans) to suppress correlated $1/f$ noise via baseline
polynomial subtraction and to remove artifacts such as `glitches'
caused by cosmic rays.  Our maps are astrometrically aligned with
pre-existing radio and {\it Spitzer}/MIPS 24\,\mum\ data to better
than 1\arcsec\ accuracy using stacking.

Because SPIRE is diffraction limited, the beam size varies
considerably over the three bands from 17.6\arcsec\ at 250\,\mum\ to
35.3\arcsec\ at 500\,\mum\ (FWHM).  In the absence of confusion noise,
and for instrument noise that is as white as in the SHIM maps, the
mathematically optimum procedure for point-source selection is to
smooth the maps by the beam.  Since our sources are, by definition,
brightest at 500\,\mum , we therefore smooth all three maps to an
identical resolution of $\sqrt{2} \times 35.3\arcsec = 49.8\arcsec$.
Since SPIRE maps have significant confusion noise, this approach is
not ideal -- in the future we plan to improve this procedure by
adopting a filtering procedure that takes confusion noise into
account, such as that described by \citet{chapin11}.

In order to facilitate this beam matching, we construct SHIM maps with
4\arcsec\ pixels at all bands.  This results in incomplete coverage, particularly
at 500\,\mum\ (where SPIRE has the lowest sampling density), which is
a problem for the convergence of the default SHIM map-maker.  However,
SHIM makes it possible to re-use baseline and glitch information from
maps made at a standard resolution, where convergence is not a
problem, and so we have done so for our analysis.  We then
Gaussian-smooth all the maps to the final resolution,
which also fills in the missing coverage due to the small pixels, and
then suppress large-scale structure by subtracting a smoothed version
of the map (using a smoothing scale of 3$^{\prime}$).  Tests on
simulated data show that these steps have no significant effect on the
accuracy of photometry from the resulting maps, although the smoothing
does increase the confusion noise.

\subsection{Object Identification in Difference Maps}
\label{subsec:diff}
\nobreak 
Prior to source identification, we form a weighted combination of the
SPIRE maps to reduce the confusion from typical, bluer SPIRE galaxies.
For simplicity -- and because experiments with more complicated
(quadratic or cubic) map combinations do not seem to provide better
performance -- we restrict ourselves to a linear combination. With no
loss in generality this can be formulated as:
\begin{equation}
D = k_1 M_{500\,\mum} + k_2 M_{350\,\mum} \pm \sqrt{1 - k_1^2 - k_2^2} M_{250\,\mum},
\end{equation} 
with $\left|k_{1}\right|$ and $\left|k_{2}\right|$ both $\le 1$.  The
term multiplying the 250\,\mum\ map is simply an arbitrary
normalization convention, since overall scale factors do not affect
our analysis.  The sign of this term is another parameter.  However --
as one would expect, because we are trying to select sources with
$\fplw \ge \fpsw$ -- the minus sign performs much better,
so we adopt it henceforth.  Because the maps considered in this paper
are confusion-noise dominated, there is little advantage to adopting
different $k$ values for the different fields (e.g., applying further
noise weighting). We optimized the values of $k_1$ and $k_2$ through
simulation, starting with the \citet[][hereafter B11]{bethermin11}
model.  This model predicts essentially no bright 500\,\mum-risers
(see \S\ref{subsec:modelcompare}), and so we also include additional
red sources in order to evaluate our efficiency at detecting them.

For this purpose we used a preliminary version of the $k$s -- which
turned out to be very similar to their final values -- and injected
fake red sources over a range of flux densities and at approximately
the observed space density into maps generated from the B11 model,
including instrument noise.  We then explored a range of values
for $k_1$ and $k_2$ by applying the source selection procedure
(described in the next section), and used this to estimate the purity
(the fraction of detected sources which are red and above our
\fplw\ cutoff) and completeness (the fraction of the injected red
sources which are recovered) for various values of the $k$'s.
Adopting the product of these two factors as our metric, we identified
a broad locus of $k_1$, $k_2$ values which provide similar
performance.

In order to down-select to the final set of parameters from this set,
we further simplified by choosing $k_2 = 0$ and then adopting the value
of $k_1$ from the high-performance locus that minimized the variance
in the $D$ map for our deepest field (GOODS-N).  Our final values are
$k_1 = 0.92$, $k_2 = 0.0$.  Therefore, for the rest of this analysis,
we work with the specific difference map
\begin{equation}
 D = 0.920 M_{500\,\mum} - 0.392 M_{250\,\mum}.
\end{equation} 
It is somewhat surprising that including the 350\mum\ map in the
combination did not improve the performance of the source finding, but
this set of parameters performs as well as any other set of parameters
in our tests.  Any modifications to our procedure (such as
changing the applied smoothing, or applying this procedure to
shallower, instrument noise dominated maps) are likely to change the
optimum values of the $k$s.

We use the measured instrument noise properties of the input SHIM maps
to estimate the instrument noise in the difference map ($D$),
including the effects of the smoothing, but neglecting the
correlations that it imposes between neighboring pixels.  We measure
the confusion noise in the $D$ maps following the approach of
\citet{ngu10}, and find a confusion noise of $\sigma_{\mathrm{conf}} =
4.2\,\mathrm{mJy}$.  This is less than the confusion noise in the {\it
  un}-smoothed single-band SPIRE maps ($\sim 6$ mJy), despite the fact
that we have degraded the resolution significantly by smoothing; this,
in some sense, is the point of forming the difference map -- it
effectively removes, or at least de-weights, the bulk of the (non-red)
SPIRE-detected population, while reducing the signal from red sources
by considerably less.  Note that the slope of the mean relation
between \fplw\ and \fpsw\ map pixels is $\sim 2.3$
(Figure~\ref{fig-sel}).  The fact that this is so close to the ratio
of our coefficients (2.36), which were derived from a process that
also included considerations related to red source recovery, is
encouraging.

The noise in areas of the maps without cross scans is more complicated
and difficult to simulate.  Therefore, we masked the edges of our maps
to exclude such regions as well as to provide better uniformity in
sampling density.  We then search for sources in the difference map
using
StarFINDER\footnote{http://www.bo.astro.it/StarFinder/paper6.htm}
\citep{diolaiti00}, a package designed for source detection in crowded
fields.  Because the instrument noise estimate ignores the
correlations induced by the smoothing procedure, and because the
confusion noise distribution is highly non-Gaussian, it is important
to use simulations to select the minimum signal-to-noise (S/N) requirement to
impose on the source finding in order to optimize the purity and
completeness of the resulting catalog.  Based on such simulations (see
\S \ref{subsec:sourcedens}), we have adopted a minimum S/N of 4, where
the noise is the quadrature sum of the 1$\sigma$ instrument and
confusion noise.  Because our fields range considerably in depth, the
S/N requirement translates to different limits in $D$ for different
fields.  For the large, shallow fields that dominate our
catalog, $D > 25.0$\,mJy in FLS, and $D > 23.4$\,mJy in Lockman-SWIRE.  The
depths for the smaller, deeper fields are $D > 19.4$\,mJy in
Lockman-North and Lockman-East, and $D > 17.3$\,mJy in GOODS-N.

\subsection{Selection of Sources Rising at 500\,\mum}
\label{sec-sel}
\begin{figure}
\epsscale{0.9}
\plotone{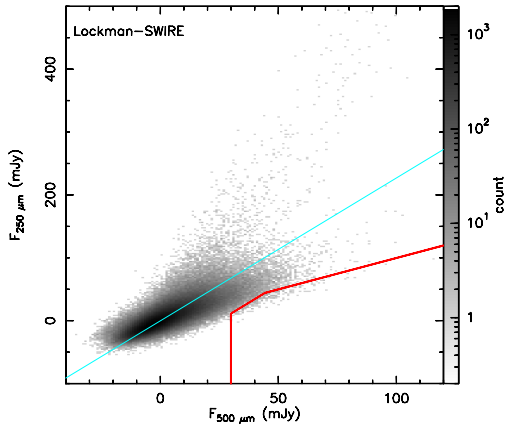}
\caption{Two-dimensional histogram illustrating the difference map selection
  used in this paper.  The axes are SPIRE 500\,\mum\ and 
  250\,\mum\ map-based flux densities.  The grayscale gives the number of
  counts in each flux-flux bin for the Lockman-SWIRE field.  The cyan line
  shows the ``average'' color of the galaxies, based on minimizing the
  variance in the difference image $D(k_1)$.  The red line segments show
  the boundary for object selection, a combination of the threshold
  in $D$, the requirement $\fplw \geq \fpsw$, and
  the $\fplw \ge 30$ mJy requirement.  The further
  selection criteria $\fplw \geq \fpmw$ and $\fpmw \geq \fpsw$ are not
  illustrated here.  The input data are the oversampled images; a
  single galaxy occupies multiple pixels and hence multiple bins.
  This potential redundancy in object selection is eliminated by the
  peak finding algorithm.  Note that the maps have zero mean before and
  after point-source filtering, and the bins with significant negative
  flux densities represent pixels between objects.}
\label{fig-sel}
\epsscale{1}
\end{figure}

\nobreak 
The choice $k_1 = 0.92, k_2 = 0$ in the difference images allows
``leakage'' of bright sources which are somewhat blue ($1 < \fpsw /
\fplw < 2.3$).  Selecting only sources which rise in $500\,\mum$ band
thus requires an additional selection step.  At the position of each
detected source in the difference maps, we measure the three SPIRE
flux densities from the smoothed maps, since, given our typical
positional uncertainties (\S\ref{subsec:interfer}), measuring the flux
densities from the un-smoothed maps results in significant biases.  We
use these measurements to further impose the requirements $\fplw \geq
\fpmw \geq \fpsw$ and $\fplw \geq 30$ mJy, the effects of which are
shown in Figure~\ref{fig-sel}.  The last criterion selects sources
which are bright enough for relatively easy follow-up.  Furthermore,
sources fainter than this limit generally have very low detection
efficiencies in the SPIRE data using our method (typically 5--20\%),
and the uncertainties in those efficiencies are quite large because we
do not know their flux densities precisely.  Therefore, if we were to
include sources below this limit, it would significantly degrade the
precision of our source density measurement (see \S\ref{sec:numdens}).

Once sources are identified from the difference maps, we match our
targets against the HerMES SCAT catalogs \citep{smi11}, which are
based on the un-smoothed maps, and provide more precise flux
measurements; recall that the reason for not using the catalogs to
select red targets is because a catalog search results in a
larger number of false sources, but once red targets are identified by
the map search this concern no longer applies.  This is not possible
in all cases, because some of our red sources were not detected by
SCAT, possibly because they are too faint in the bluer (250 and
350\,\mum) SPIRE bands.  We use the SCAT v21 catalogs selected at
350\,\mum\ for this purpose -- the 250 and 500\,\mum\ selected
catalogs (where available) are generally too incomplete for red
sources for our purposes.

In addition, we have visually inspected all sources which pass these
cuts in order to remove (rare) noise artifacts, and to note blends of
faint sources.  The more difficult cases involve noise artifacts on top
of real sources which cause them to appear red.  However, in general, 
such false sources are quite easy to detect in the
un-smoothed maps, and have been removed from our analysis.
In \S\ref{subsec:purity} we make use of simulations to quantify the
effects of instrument noise scattering ``orange'' sources across
the selection boundary.

\begin{figure*}[t!]
\includegraphics[width=1\textwidth]{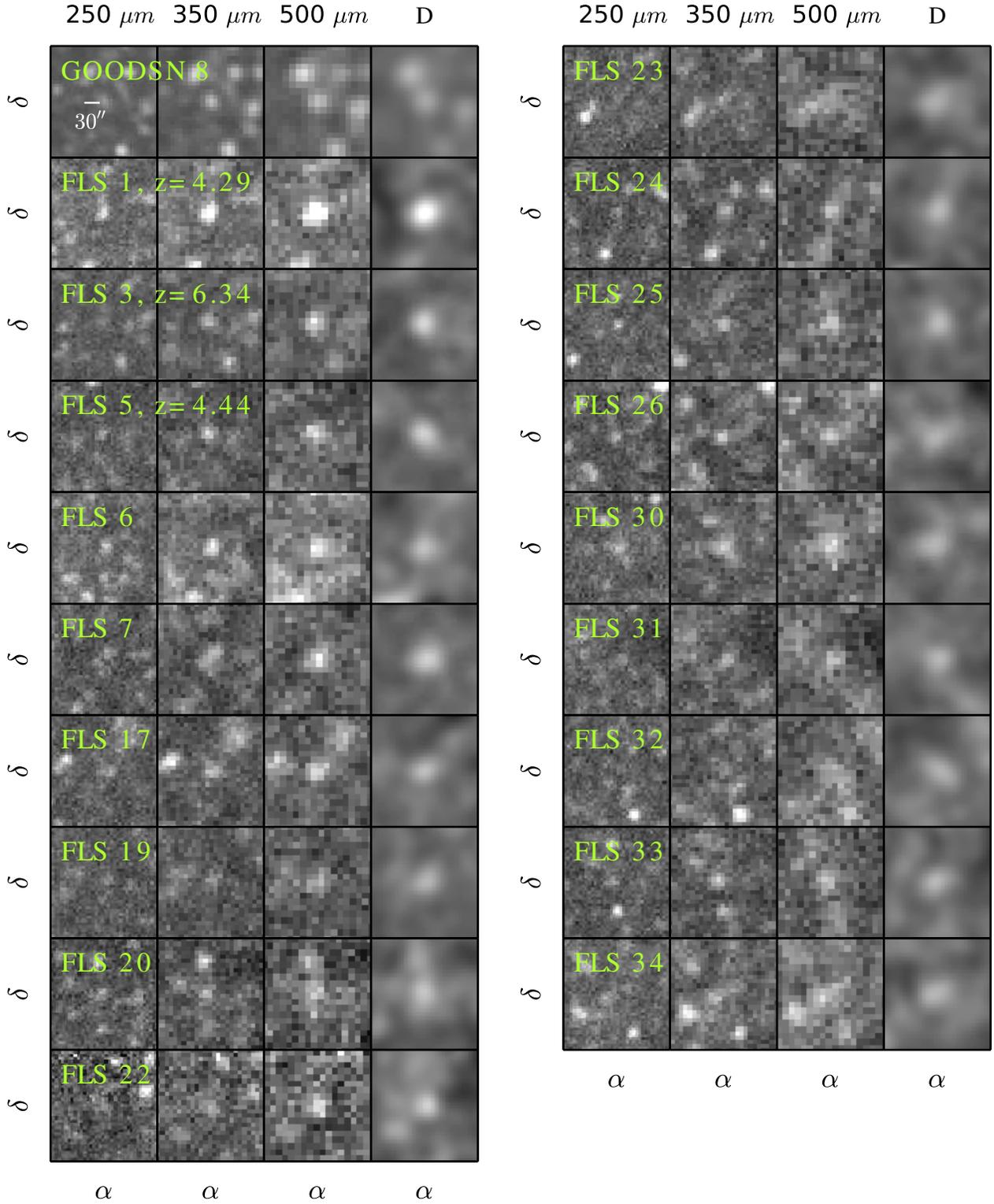}
\caption{Postage stamps of all of the GOODS-N and FLS sources in our
  catalog.  The grayscale range for each cutout is -35 to +50 mJy.
  Sources with secure spectroscopic redshifts are noted.  For explanation of
  the `D' image, see \S\ref{subsec:diff}. \label{fig:flscutouts}}
\end{figure*}

\begin{figure*}[t!]
\includegraphics[width=1\textwidth]{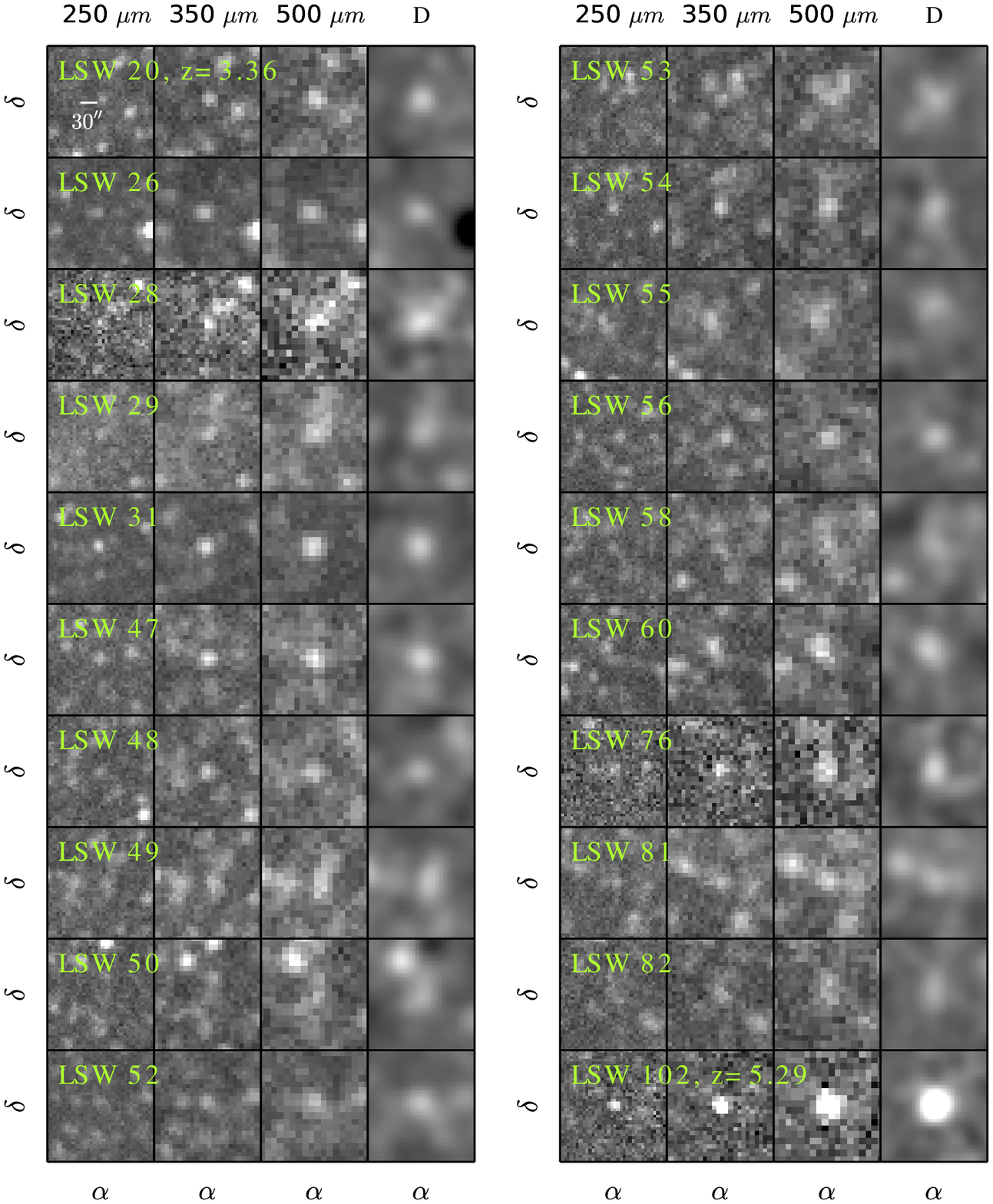}
\caption{Postage stamps of all of the LSW sources in our catalog.  The
  grayscale range for each cutout is -35 to +50 mJy. Sources with
  Sources with secure spectroscopic redshifts are noted.  For explanation of
  the `D' image, see \S\ref{subsec:diff}. \label{fig:lswcutouts}}
\end{figure*}

The 38 sources which meet all selection criteria are listed in
Table~\ref{tbl-src} and form the basic sample for this paper.  There
is one in GOODS-N, 18 in FLS, and 19 in the combined Lockman regions.
In addition we include in the table two bright sources (LSW~28 and
LSW~102) from outside our field mask (due to the lack of cross-scans
at their positions), but do not include them in our formal catalog.
Both sources easily pass our selection criteria, and are clearly real;
in fact the redshift of LSW~102 is 5.3 (Perez-Fournon et al.\ , in
preparation).  LSW~28 is a few arcminutes away from the 2nd magnitude
star $\beta$ Ursa Major, which causes complications for
optical/near-IR follow-up.  For comparison, we also include in the
table (but not in the statistical analysis) the known $z = 4$ source
pair GN20/20.2 \citep{pop05,dad09b}; it barely fails our selection
criteria.  Note that this pair has a separation of 24\arcsec, and
hence is resolved in the 250\mum\ channel.  The quoted SCAT and radio
fluxes are the sum of both.  The positions in Table~\ref{tbl-src},
and throughout this paper, are the centroids in the difference
($D$) images and have $\sim$5\arcsec\ 1$\sigma$ uncertainty (see
\S\ref{subsec:interfer}).  Sources identified as blends or which do
not satisfy $\fplw \ge \fpmw$ are identified in the table.  For
comparison, the SCAT catalogs of these fields contain $\sim 1400$
non-red sources with $\fplw > 30$ mJy.  Based on their flux densities,
pre-existing lensing models for the DSFG population predict that most
of our sources should not be significantly lensed \citep[e.g.,][]{
  negrello07, wardlow:13}.  However, these models were developed for
lower-$z$ DSFGs, so this will have to be tested against future
interferometric observations. Cutouts of all of the sources in our
final catalog are shown in Figures~\ref{fig:flscutouts} and
\ref{fig:lswcutouts}.

\begin{table*}[t!]
  \centering
  \caption{SPIRE and radio fluxes for sources selected as 
    described in Section~\ref{sec-sel}.}
  \begin{tabular}{l | ll | llll | lll | l}
  \hline
  Source & \multicolumn{2}{c}{Position} &  
 \multicolumn{4}{c}{Smoothed Map} & \multicolumn{3}{c}{SCAT} & Radio \\
 Name & $\alpha_{2000}$ & $\delta_{2000}$ & $S_{250\,\mum}$ & 
 $S_{350\,\mum}$ & $S_{500\,\mum}$ & $D$ & $S_{250\,\mum}$ & $S_{350\,\mum}$ & 
 $S_{500\,\mum}$ & $S_{21\,cm}$ \\
   & (deg) & (degrees) & (mJy) & (mJy) & (mJy) & (mJy) & (mJy) & 
 (mJy) & (mJy) & ($\mu$Jy) \\
\hline\hline
GOODSN 8$^{x}$ & 188.964  &  62.363  & 22.7 & 33.6 & 35.9 & 20.9 & \nodata & \nodata & \nodata & $27 \pm 4$ (\S\ref{subsec:radio}) \\
\hline
FLS 1            & 257.072  &  58.479  & 50.7 & 82.9 & 87.4 & 61.6 & 62.3 & 94.0 & 98.2 & $<840$ \\
FLS 3            & 256.697  &  58.772  & 13.4 & 26.8 & 47.5 & 37.9 & 12.0 & 32.4 & 47.3 & $59 \pm 11$ \\
FLS 5            & 260.204  &  59.773  & 17.1 & 40.0 & 47.7 & 37.3 & 24.7 & 45.1 & 48.2 & $<920$ \\
FLS 6            & 256.797  &  60.470  & 39.3 & 53.0 & 53.7 & 26.6 & \nodata & \nodata & \nodata & $<1200$ \\
FLS 7$^{b,x}$     & 257.511  &  59.089  & 24.2 & 50.2 & 51.7 & 40.4 & \nodata & \nodata & \nodata & $<1030$ \\
FLS 17           & 260.794  &  60.331  & 33.7 & 43.0 & 48.7 & 27.7 & 28.4 & 42.5 & 48.6 & $<960$ \\
FLS 19           & 258.754  &  60.429  & 11.5 & 21.9 & 36.9 & 27.2 & 15.0 & 21.2 & 31.4 & $<980$ \\
FLS 20           & 256.407  &  60.370  & 29.9 & 39.5 & 50.8 & 36.4 & 25.1 & 34.7 & 43.5 & $<1080$ \\
FLS 22           & 257.863  &  60.632  &  2.5 & 31.9 & 40.0 & 34.3 & 14.5 & 33.2 & 37.1 & $<1010$ \\
FLS 23$^b$      & 260.334  &  58.571  & 29.6 & 39.6 & 41.6 & 32.9 & 12.3 & 26.7 & 31.7 & $<1010$ \\
FLS 24           & 261.196  &  59.556  &  2.0 & 26.1 & 32.6 & 31.7 & 16.3 & 32.2 & 34.3 & $<960$ \\
FLS 25           & 259.788  &  60.121  & 29.6 & 41.2 & 45.6 & 32.6 & 31.6 & 40.3 & 42.9 & $<1020$ \\
FLS 26$^x$       & 257.798  &  58.963  & 15.8 & 30.8 & 35.0 & 30.8 & \nodata & \nodata & \nodata & $<980$ \\
FLS 30$^x$       & 258.364  &  58.401  & 47.5 & 52.3 & 54.7 & 29.7 & \nodata & \nodata & \nodata & $<910$ \\
FLS 31$^x$       & 257.644  &  58.639  & $-2.0$ & 26.4 & 33.0 & 27.6 & \nodata & \nodata & \nodata & $<990$ \\
FLS 32$^x$       & 261.148  &  60.053  & 10.9 & 20.6 & 34.2 & 26.4 & \nodata & \nodata & \nodata & $<1000$ \\
FLS 33           & 259.879  &  60.475  & 26.3 & 42.1 & 44.3 & 26.9 & 21.6 & 39.6 & 40.6 & $<1010$ \\
FLS 34$^x$       & 257.110  &  59.680  & 16.7 & 33.7 & 40.7 & 27.8 & 16.7 & 33.7 & 40.7 & $<950$ \\
\hline
LSW 20           & 162.885  &  56.605  &  4.5 & 21.5 & 39.0 & 33.1 & 17.6 & 36.6 & 43.9 & $<920$ \\
LSW 25$^r$       & 159.435  &  57.198  & 47.5 & 59.9 & 64.3 & 41.6 & 51.3 & 66.9 & 69.7 & 127,000 \\
LSW 26$^y$       & 161.056  &  58.770  & 13.4 & 30.3 & 31.4 & 27.4 & 23.9 & 39.0 & 34.2 & $<980$ \\
LSW 29           & 161.935  &  57.942  & 30.1 & 50.0 & 51.2 & 31.5 & 32.5 & 47.5 & 50.3 & $<980$ \\
LSW 31$^x$       & 160.973  &  59.361  & 40.6 & 53.8 & 57.2 & 39.8 & \nodata & \nodata & \nodata & $<970$ \\
LSW 47$^x$       & 162.148  &  57.936  & 29.0 & 52.5 & 53.5 & 38.7 & \nodata & \nodata & \nodata & $<960$ \\
LSW 48$^y$       & 163.427  &  56.590  &  3.4 & 22.1 & 29.8 & 34.1 & 22.6 & 41.7 & 40.9 & $<950$ \\
LSW 49           & 159.652  &  57.698  & 27.2 & 44.4 & 48.1 & 33.9 & 23.1 & 38.5 & 40.1 & $<1040$ \\
LSW 50$^x$       & 163.274  &  56.684  & $-$2.1 & 18.0 & 30.3 & 32.7 & \nodata & \nodata & \nodata & $<950$ \\
LSW 52           & 163.954  &  57.566  & 17.3 & 32.6 & 43.7 & 31.6 & 16.3 & 33.0 & 40.2 & $<990$ \\
LSW 53           & 163.135  &  58.607  & 31.5 & 40.7 & 44.9 & 31.9 & 15.8 & 29.7 & 32.6 & $<1030$ \\
LSW 54           & 163.743  &  57.061  & 31.9 & 41.0 & 44.0 & 30.9 & 36.6 & 46.0 & 46.7 & $<920$ \\
LSW 55$^x$       & 164.774  &  57.751  &  7.7 & 24.7 & 31.2 & 29.2 & \nodata & \nodata & \nodata & $<970$ \\
LSW 56           & 165.086  &  58.028  & 23.6 & 41.1 & 43.3 & 29.5 & 28.4 & 41.7 & 44.8 & $<1020$ \\
LSW 58           & 163.391  &  56.608  &  7.2 & 28.8 & 32.3 & 27.8 & 15.6 & 29.5 & 34.1 & $<950$ \\
LSW 60$^x$       & 161.914  &  60.087  & 28.6 & 47.3 & 47.6 & 32.9 & \nodata & \nodata & \nodata & $<1210$ \\
LSW 76$^x$       & 165.353  &  59.063  & 41.1 & 60.1 & 61.3 & 40.4 & \nodata & \nodata & \nodata & $<1030$ \\
LSW 81$^y$       & 163.649  &  56.522  & 20.2 & 38.7 & 41.4 & 26.6 & 29.6 & 45.0 & 43.1 & $<1010$ \\
LSW 82           & 165.420  &  58.227  & 21.9 & 33.4 & 37.5 & 25.9 & 16.4 & 34.7 & 37.2 & $<920$ \\
\hline
\hline
LSW 28$^c$     & 165.366  &  56.326  & 35.1 & 53.6  & 64.0 & 45.2 & 33.4 & 55.9 & 60.0 & $<960$ \\
LSW 102$^c$    & 160.211  &  56.115  & 51.0 & 122.3 & 140.0 & 109.4 & 49.7 & 118.1 & 140.4 & $<1010$ \\
\hline
GN20/20.2$^{b,o}$   & 189.300  &  62.370  & 52.3 & 68.6 & 65.4 & 38.2 & 59.4 & 75.2 & 74.1 & $289\pm13$ \\
\hline
\end{tabular}
\begin{flushleft}
  $^b$Source is clearly detected as blended in un-smoothed 250\,\mum\ maps.\\
  $^c$Source is in a region of the maps that does not have redundant scans, 
   and hence not included in the statistical analysis.\\
  $^o$Known $z=4$ source not detected by our pipeline, but included for 
   comparison.\\
  $^r$Radio-loud object dominated by synchrotron emission; omitted 
      from sample.\\
  $^x$Source not present in SCAT v21 350\,\mum\ selected catalog.\\
  $^y$Source with $\fpmw > \fplw$ in the SCAT catalog.
\end{flushleft}
\tablecomments{
  SPIRE and radio parameters for sources detected via our
  map-based method.  The first three 250/350/500\,\mum\ flux densities
  are those derived from the smoothed maps, and $D$ is the flux
  density measured in the linear map combination described in the
  text.  The SCAT-derived SPIRE flux densities are measured from the
  un-smoothed maps at the native SPIRE resolution, and are from the
  SCAT v21 HerMES catalogs. The typical instrumental noise uncertainty
  in both sets of SPIRE flux densities is 2--3 mJy.  The calibration
  uncertainties for SPIRE are about 5\%, and the confusion noise is
  $\sim 6$ mJy in the un-smoothed maps; note, however, that both are
  highly correlated between bands, so have a minor effect on the color
  ratios between SPIRE bands.  The SCAT flux densities, where
  available, are expected to be more accurate.  The $\sim 1000\,\mu$Jy 
  21\,cm flux density upper limits are from the FIRST or NVSS catalogs.  For
  brighter sources, the positional accuracy is $\sim 5$\arcsec\ (see
  \S\ref{subsec:interfer}).  The names of these sources are drawn from
  a master list that includes additional, non-red HerMES sources. \\
\label{tbl-src}}
\end{table*}


\subsection{Screening against Synchrotron Sources}
\nobreak 
Since bright synchrotron-emitting galaxies may have far-IR spectra
consistent with our selection criteria at modest ($\le 1$) redshifts,
we screen against such objects by searching for bright ($>$1 mJy)
radio counterparts.  We used the NVSS and FIRST 21\,cm catalogs
\citep{bec95, con98}, which cover our fields, and found one such
radio-bright source coincident with a SPIRE source in our sample:
LSW~25.  This turns out to be a known BL Lac at a redshift of $z=0.83$
\citep{rich09}, and we exclude it from all further analysis.  Another
SPIRE source, LSW~48, lies 21\arcsec\ away from a radio source with a
21\,cm flux density of 2.8\,mJy; given the typical positional
accuracy, this is unlikely to be due to the same source, so we retain
LSW~48 in our sample.  FLS~3 and GOODSN~8 have significantly deeper
21\,cm observations (see \S\ref{subsec:radio}).  For the remaining
sources we provide the catalog detection limit at that position in
Table~\ref{tbl-src}.  Four more of our targets are covered by deeper,
narrower observations at 19--20\,cm, which are presented in
\S\ref{subsec:radio}.  Other than LSW~25, the radio flux densities of
all our sources are faint enough to be consistent with the FIR/radio
correlation seen for star-forming galaxies at $z \simeq 4$
\citep{ivison:10}, implying that they are dust-emission dominated.


\section{Basic Observed Far-IR Source Properties}
\label{sec:basic}
\nobreak 
For each source, we fit a modified blackbody spectrum to the three
SPIRE flux densities, with the primary goal of determining the
wavelength at which the spectrum reaches its maximum.  This
information is used to estimate the redshift distribution of the
population in \S\ref{subsec:pz}.  Given the small number of SPIRE
bands, it is not possible to fully constrain such a model.  In such
cases it is common in the literature to assume that the emission is
optically thin and furthermore to fix the log-slope $\beta$ of the
optical depth curve to some nominal value ($\approx 1.5$): $S_{\nu}
\left(\lambda_{\mathrm{obs}}\right) \propto \lambda_{obs}^{-\beta}
B_{\nu} \left( T_{\mathrm{d}} /\left(1+z\right)\right)$, where
$B_{\nu}$ is the Planck function, $T_{\mathrm{d}}$ the dust
temperature, and $\lambda_{\mathrm{obs}}$ the observer frame
wavelength.  (Note that, as is common in sub-mm/mm astronomy, we work
in $S_{\nu}$ units but expressed as a function of wavelength.)

In this paper we instead use the slightly more general formalism of an
optically-thick modified blackbody:
\begin{eqnarray}
S_{\nu} \left(\lambda_{\mathrm{obs}}\right) = \Omega & \times & \left( 1 -
\exp\left[-\left(\lambda_0\left( 1+z \right) / 
\lambda\right)^{\beta}\right]\right) \times \nonumber \\
&\qquad& B_{\nu}\left(\lambda_{\mathrm{obs}}; T_{\mathrm{d}} / \left( 1+z 
 \right) \right),
\label{eqn:mbb}
\end{eqnarray}
where $\lambda_0$ is the (rest-frame) wavelength where the optical
depth reaches unity and $\Omega$ is the solid angle subtended by the
source.  To the blue side of this relation we attach a power law
$S_{\nu} \propto \lambda^{\alpha}$, joining the two by requiring the
SED and first derivative to be continuous.  We fit this model to
the data using an affine-invariant Markov Chain Monte-Carlo (MCMC)
approach, as discussed in the Appendix.  In the current section,
where we only analyze SPIRE photometry, we marginalize over broad
priors in the less well constrained model parameters rather than
fixing them.  Later, for sources where more photometry is available
(\S\ref{subsec:detailed-fits}) we remove these priors.  In all cases
in this paper, $\alpha$ is poorly constrained, with the fits only
providing a lower limit.  Because of the way the blue-side power law
is joined to the modified blackbody portion of the SED (requiring that
the first derivatives match), this simply amounts to a statement that
the merge point between the power law and the thermal SED lies at
wavelengths lower than the shortest wavelength photometry point
(250\,\mum).  That is, in all cases, the quality of our fits is not
improved by the addition of the Wien-side power law, and, while
formally included, $\alpha$ is effectively not a parameter of our
fits.

Due to the poor constraints possible with only SPIRE data, we
marginalize over broad Gaussian priors in $\beta$ and $\lambda_0
\left(1+z\right)$.  There is relatively little guidance in the
literature for the appropriate value for $\lambda_0$, although
100\,\mum\ has been adopted in some studies.  In general, both {\it
  Herschel} and longer $\lambda$ observations are required to
constrain $\lambda_0$, and hence it has been measured for relatively
few sources.  In order to determine an appropriate prior, we have
applied this same model to literature sources with a sufficient
quantity of high quality photometry such as Arp~220
\citep{Rangwala:2011}, LSW~1 \citep{Conley:11}, the Cosmic Eyelash
\citep{Swinbank:10}, FLS~3 \citep{rie13}, XMM~1 \citep{fu:13}, ID 141
\citep{cox11}, as well as three of the red sources presented in this
paper (\S\ref{subsec:detailed-fits}).  We conclude that, with the
exception of FLS~3, all sources are significantly better fit by an
optically thick model rather than an optically thin one, with $\Delta
\chi^2$ values of $>6$, and that $\lambda_0$ ranges from 190 to
270\,\mum\ (rest frame).  If our sources have a typical redshift of
$4.5 \pm 1$, this corresponds to a Gaussian prior on $\lambda_0 \left(
1+z \right)$ of $1100 \pm 400$\,\mum, which we adopt (truncating the
prior to exclude negative values).  We further assume a Gaussian prior
on $\beta$ of $1.8 \pm 0.3$ \citep{draine06}.  $\beta$ is typically
believed to be below 2, but examples of larger values are known
\citep[e.g.,][]{schwartz:82}, so we do not set a hard upper limit in
our analysis.

We use the SCAT~v21 catalog flux densities where available, and
otherwise the smoothed map fluxes (Table~\ref{tbl-src}).  In addition
to the instrumental uncertainties in the catalog fluxes (typically
2--3\,mJy), we include 4\% grayscale calibration uncertainty, a 1.5\%
per-band uncorrelated calibration uncertainty \citep{Bendo:13}, and
confusion noise due to the modest SPIRE resolution.  We estimate the
covariance matrix of the latter using simulations based on the B11
model, (slightly) scaled to match the observed confusion noise
\citep{ngu10}.  Note that we use a different confusion noise
covariance matrix for fluxes derived from the un-smoothed and smoothed
maps (SCAT vs.\ map based) -- in the latter case the confusion noise
is larger.  However, in either case, as the confusion noise and
calibration uncertainties are highly correlated among the bands, they
have relatively little effect on estimates of the peak wavelength and
temperature.  It is clear that longer-wavelength observations
($\lambda$ = 750--2000\,\mum) would significantly improve our SED
constraints.

\begin{figure}
\epsscale{1.1}
\plotone{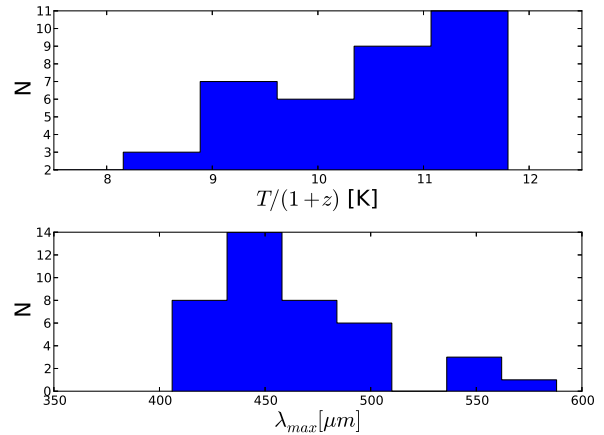}
\caption{Results of the modified blackbody fits to the SPIRE
  photometry of our sources.  The top panel shows the measured
  $T_{\mathrm{d}}/\left(1+z\right)$ distribution, and the bottom panel
  the observer-frame wavelength at which $S_\nu$ peaks.
\label{fig:mbb_hist}}
\end{figure}

The fit results are given in Table~\ref{tbl-otherflux}.  For most of
our sources, the SPIRE data are well fit by a single temperature
modified blackbody, with the possible exception of FLS~31.  Histograms
of $T_{\mathrm{d}}/\left(1+z\right)$ and $\lambda_{\mathrm{max}}$ (the
observer frame peak wavelength of $S_{\nu}$ estimated from our SED
fits)\footnote{Note that in an optically thick modified blackbody
  model $\lambda_{\mathrm{max}} \not\propto
  \left(1+z\right)/T_{\mathrm{d}}$, even if all other parameters are
  fixed.} are shown in Figure~\ref{fig:mbb_hist}.  These central
values are relatively insensitive to the exact form of the priors,
although $\lambda_{max}$ is less sensitive than
$T_{\mathrm{d}}/\left(1+z\right)$.  For example, if we change the
$\beta$ prior to $1.5 \pm 0.3$, $\lambda_{\mathrm{max}}$ changes by
$<1\%$ and $T_{\mathrm{d}}/\left(1+z\right)$ increases by about 5\%.
$L_{\mathrm{IR}}$, however, is fairly sensitive to the $\beta$ prior
(in the absence of additional photometry and assuming a fixed $z$).
The results are even less sensitive to the $\lambda_0$ prior,
particularly $\lambda_{\mathrm{max}}$, although if we adopt an
optically-thin model ($\lambda_0 \rightarrow 0$),
$T_{\mathrm{d}}/\left(1+z\right)$ drops by about 15\%.

FLS~3 ($z=6.3$) is one of the reddest sources, but there are others as
red in our sample.  Some of the sources have mm-wave follow-up
(see \S\ref{sec-followup}).  We can use this to check whether our priors
were reasonable by using our model to predict these flux densities and
comparing them with observations, and by comparing the predicted
$\lambda_{\mathrm{max}}$ from SPIRE data to that measured when the peak is
better constrained by adding mm-data.  With the exception of FLS~17,
which is significantly fainter than predicted at 1.3mm, we find that
this model does a good job predicting the mm-fluxes and
$\lambda_{\mathrm{max}}$, albeit with large (10-25\%) uncertainties.  For
example, for FLS~5 $\lambda_{\mathrm{max}}\left(\mathrm{SPIRE}\right) = 452 \pm
37\,\mum$, while the value measured with the addition of mm data is
$445 \pm 20\,\mum$.  For FLS~3 the values are $550 \pm 86\,\mum$ and
$537 \pm 22\,\mum$; the modest improvement in the
precision on $\lambda_{\mathrm{max}}$ when mm-data are added is because of the
additional priors applied to the SPIRE-only fits.

\begin{table*}[t!]
 \centering
 \caption{Quantities derived from the SPIRE observations}
 \begin{tabular}{l|cccc|c}
  \hline
  Source & $T_{\mathrm{d}}/\left(1+z\right)$ & $\lambda_{\mathrm{max}}$ & 
  $L_{\mathrm{IR}}\left(z=4.7\right)^a$ &  $\chi^2$ & $\epsilon^b$ \\
   & (K) & (\mum ) & ($10^{12}\,\mathrm{L}_{\odot}$) & \\
\hline\hline
GOODSN 8         & $11.3 \pm 3.0$ & $433 \pm 83$   & $37 \pm 20$ & 0.08 & $0.51 \pm 0.17$  \\
\hline
FLS 1            & $10.8 \pm 0.8$ & $440 \pm 21$   & \S\ref{subsec:detailed-fits} & 0.02 & $0.68 \pm 0.08$  \\
FLS 3            & $8.1 \pm 1.2$  & $550 \pm 56$   & \citet{rie13} & 0.02 & $0.88 \pm 0.04$  \\
FLS 5            & $10.3 \pm 1.2$ & $452 \pm 37$   & \S\ref{subsec:detailed-fits} & 0.12 & $0.57 \pm 0.08$  \\
FLS 6            & $11.4 \pm 2.6$ & $423 \pm 68$   & $57 \pm 26$ & 0.09 & $0.57 \pm 0.12$  \\
FLS 17           & $9.9 \pm 1.9$  & $466 \pm 50$   & $40 \pm 15$ & 0.05 & $0.63 \pm 0.13$  \\
FLS 19           & $7.2 \pm 3.3$  & $558 \pm 117$  & $24^{+14}_{-11}$ & 0.05 & $0.62 \pm 0.15$  \\
FLS 20           & $8.3 \pm 3.5$  & $507 \pm 87$   & $44^{+25}_{-21}$ & 0.08 & $0.58 \pm 0.18$  \\
FLS 22           & $9.6 \pm 1.4$  & $479 \pm 49$   & $24 \pm 7$ & 0.45 & $0.50 \pm 0.18$  \\
FLS 24           & $9.9 \pm 1.6$  & $463 \pm 53$   & $24 \pm 7$ & 0.21 & $0.51 \pm 0.17$  \\
FLS 25           & $10.6 \pm 3.0$ & $436 \pm 60$   & $50^{+31}_{-23}$ & 0.15 & $0.34 \pm 0.11$  \\
FLS 26           & $11.1 \pm 2.6$ & $437 \pm 75$   & $37 \pm 18$ & 0.02 & $0.47 \pm 0.13$  \\
FLS 30           & $9.6 \pm 3.8$  & $475 \pm 108$  & $69^{+54}_{-42}$ & 0.11 & $0.48 \pm 0.15$  \\
FLS 31           & $10.7 \pm 2.3$ & $442 \pm 68$   & $30^{+11}_{-13}$ & 1.88 & $0.56 \pm 0.13$  \\
FLS 32           & $6.7 \pm 3.0$  & $588 \pm 122$  & $28 \pm 13$ & 0.07 & $0.52 \pm 0.12$  \\
FLS 33           & $10.5 \pm 1.0$ & $444 \pm 43$   & $34 \pm 9$ & 0.01 & $0.43 \pm 0.13$  \\
FLS 34           & $9.2 \pm 1.5$  & $494 \pm 55$   & $26 \pm 8$ & 0.02 & $0.60 \pm 0.09$  \\
\hline
LSW 20           & $9.3 \pm 1.3$  & $490 \pm 49$   & \S\ref{subsec:detailed-fits} & 0.03 & $0.68 \pm 0.08$  \\
LSW 26           & $11.5 \pm 1.7$ & $406 \pm 40$   & $30^{+7}_{-10}$ & 0.64 & $0.38 \pm 0.12$  \\
LSW 29           & $11.1 \pm 1.4$ & $432 \pm 37$   & $43 \pm 13$ & 0.12 & $0.59 \pm 0.09$  \\
LSW 31           & $10.6 \pm 2.5$ & $450 \pm 73$   & $57 \pm 28$ & 0.04 & $0.61 \pm 0.09$  \\
LSW 47           & $11.6 \pm 2.0$ & $412 \pm 50$   & $49^{+15}_{-19}$ & 0.21 & $0.53 \pm 0.12$  \\
LSW 48           & $10.7 \pm 1.4$ & $434 \pm 38$   & $31 \pm 8$ & 0.53 & $0.37 \pm 0.09$  \\
LSW 49           & $10.6 \pm 1.1$ & $440 \pm 43$   & $32 \pm 10$ & 0.02 & $0.45 \pm 0.15$  \\
LSW 50           & $8.1 \pm 2.5$  & $541 \pm 102$  & $26^{+8}_{-11}$ & 0.34 & $0.58 \pm 0.11$  \\
LSW 52           & $9.1 \pm 1.5$  & $495 \pm 54$   & $25 \pm 6$ & 0.02 & $0.59 \pm 0.09$  \\
LSW 53           & $9.7 \pm 1.8$  & $470 \pm 59$   & $23^{+6}_{-8}$ & 0.05 & $0.27 \pm 0.13$  \\
LSW 54           & $11.4 \pm 2.2$ & $413 \pm 44$   & $55^{+30}_{-22}$ & 0.16 & $0.46 \pm 0.14$  \\
LSW 55           & $10.3 \pm 3.0$ & $455 \pm 90$   & $32 \pm 15$ & 0.17 & $0.46 \pm 0.07$  \\
LSW 56           & $10.5 \pm 1.8$ & $443 \pm 46$   & $40 \pm 14$ & 0.06 & $0.62 \pm 0.04$  \\
LSW 58           & $9.5 \pm 1.8$  & $483 \pm 63$   & $24 \pm 7$ & 0.01 & $0.45 \pm 0.17$  \\
LSW 60           & $11.8 \pm 2.3$ & $410 \pm 59$   & $48^{+18}_{-21}$ & 0.05 & $0.52 \pm 0.14$  \\
LSW 76           & $11.5 \pm 2.1$ & $421 \pm 54$   & $56^{+17}_{-22}$ & 0.04 & $0.57 \pm 0.12$  \\
LSW 81           & $11.4 \pm 1.6$ & $414 \pm 35$   & $39 \pm 13$ & 0.06 & $0.42 \pm 0.10$  \\
LSW 82           & $9.8 \pm 1.4$  & $471 \pm 51$   & $26 \pm 7$ & 0.43 & $0.40 \pm 0.08$  \\
\hline
\hline
LSW 28           & $10.5 \pm 1.1$ & $448 \pm 32$   & $42 \pm 9$ & 0.02 & \nodata          \\
LSW 102          & $8.7 \pm 0.8$  & $476 \pm 18$   & $75 \pm 9$ & 1.02 & \nodata          \\
\hline
\end{tabular}
\begin{flushleft}
  $^a$Inferred 8--1000\,\mum\ (rest frame) infrared luminosity,
  assuming $z=4.7$ (\S\ref{subsec:pz}).  These values assume no lensing. \\
  $^b$Selection efficiency (see \S\ref{subsec:efficiency}).\\
\end{flushleft}
\tablecomments{
  The dust temperature, wavelength at which $S_\nu$ peaks
  (observer frame), $L_{IR}$, and $\chi^2$ values are from modified blackbody
  fits to the SPIRE photometry of our sources, as detailed in
  \S\ref{sec:basic}.  The effective number of degrees of freedom for
  the $\chi^2$ values is one.  The uncertainties are the 68.3\%
  confidence limits, which are quite non-Gaussian for some sources.
  Known blended sources (FLS 7 and FLS 23) are not included because
  their intrinsic flux densities are not known.
\label{tbl-otherflux}}
\end{table*}



\section{Follow-Up Observations}
\label{sec-followup}
We are carrying out a multi-wavelength follow-up program of these
sources.  By design, the SPIRE data only sample the blue side of the
thermal SED well, and so longer-wavelength sub-mm/mm observations are
particularly critical. They are used to determine whether our sources do, in
fact have thermal SEDs, and constrain their total IR luminosity, dust
mass, etc.  In addition, mm-wave line searches can be used to measure
redshifts and model molecular gas excitation properties, and
high-resolution interferometric observations can be used to identify
counterparts at other wavelengths and to study source morphologies
\citep[e.g.,][]{rie13}.

\begin{figure*}[t!]
 \centering
 \includegraphics[width=1.0\textwidth]{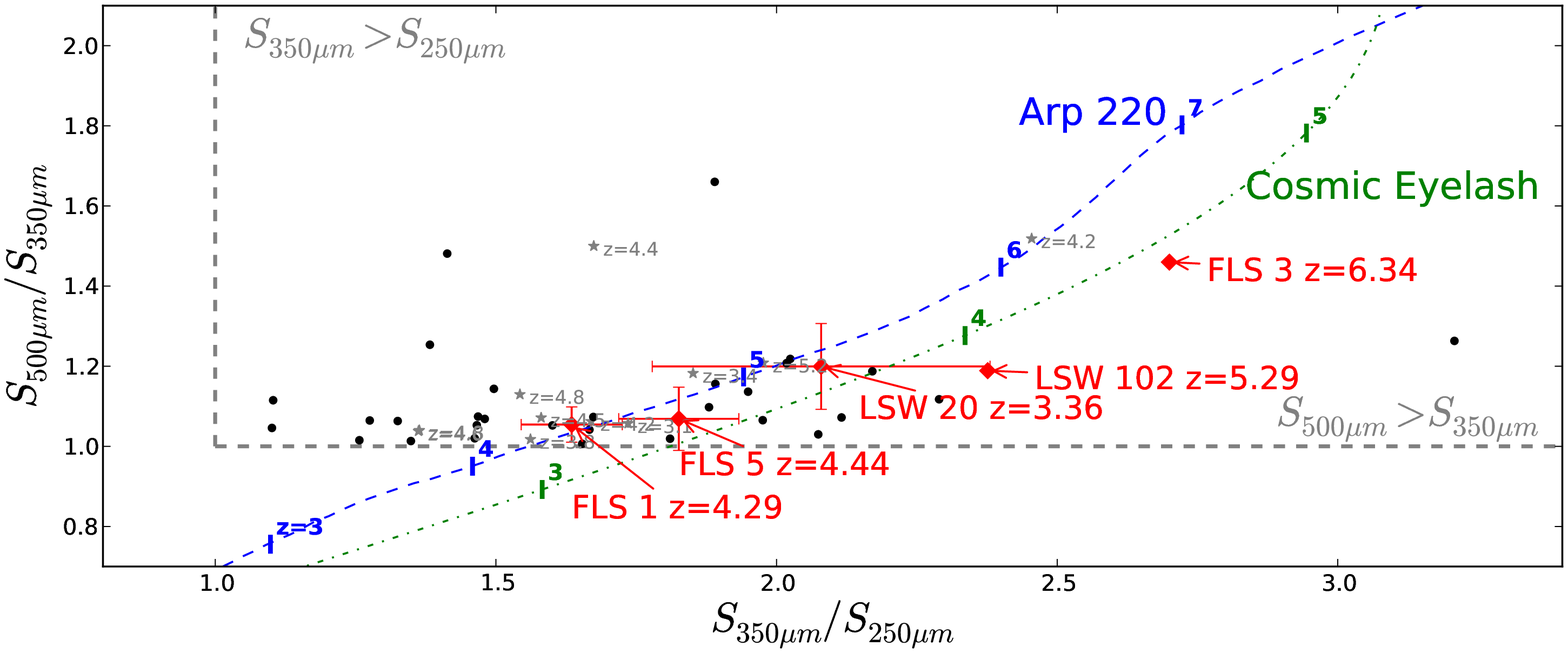}
\caption{SPIRE color ratios for sources presented in this paper (black
  dots).  The sources with redshifts are shown as red diamonds.  For
  comparison, 500\,\mum-riser sources with redshifts selected via
  other methods (i.e., not on the basis of their SPIRE colors) are
  shown as gray stars (see \S\ref{sec:otherred}).  SED tracks based on
  Arp 220 and the Cosmic Eyelash \citep{Swinbank:10} are shown for
  comparison.  They gray dashed lines represent our selection in color
  space.  Note that the uncertainties vary considerably, and are
  quite large for some of the sources.  Example uncertainties are shown
  for FLS 1, 5, and LSW 20.  \label{fig:clrplane}}
  \vskip 0.15cm
\end{figure*}

In this section we describe the relevant observations of the targets
for this paper (Table~\ref{tbl:mmflux}), concentrating on the
mm-band and radio continuum flux densities and redshifts.  Other details, such
as precise positions, morphologies, and line ratios, will be presented
elsewhere.  Four of our sources (FLS~1, FLS~3, FLS~5, and LSW~20) have
sub-mm/mm observations and known redshifts from PdBI, SMA, Z-Spec, and
CARMA, as described below. FLS~3 ($z=6.34$) has particularly extensive
follow-up, which is described elsewhere \citep{rie13}.  How the
sources with redshifts relate to the rest of the catalog sources
presented here, as well as to 500\,\mum-risers selected from the
literature by other means, is shown in Figure~\ref{fig:clrplane}.

\subsection{Pre-existing Multi-wavelength Observations}
\nobreak 
All of the HerMES fields have deep ancillary data partially covering
the SPIRE maps, and the majority of our sources have {\it Spitzer}
observations from 3.5 to 160\mum . However, counterpart identification
is found to be challenging for our galaxy sample due to the relatively
large SPIRE beam, the expected faintness of our sources at other
wavelengths compared with the depth of the available observations, and
the large number of potential counterparts in the 3.5-8\mum\ bands.
While it would be tempting to simply assume that the brightest or
closest {\it Spitzer} 24\,\mum\ source in the vicinity of each
candidate, if there is one, is the correct match, it is difficult to
justify -- or check -- this assumption without additional data.
Therefore, we postpone discussion of the near- and mid-IR counterparts
of our sources for a later publication, by which time the number of
sources with precise interferometric positions will have increased.

\subsection{Single-dish mm Observations: Z-Spec and Bolocam}
\begin{figure}
\epsscale{1.1}
\plotone{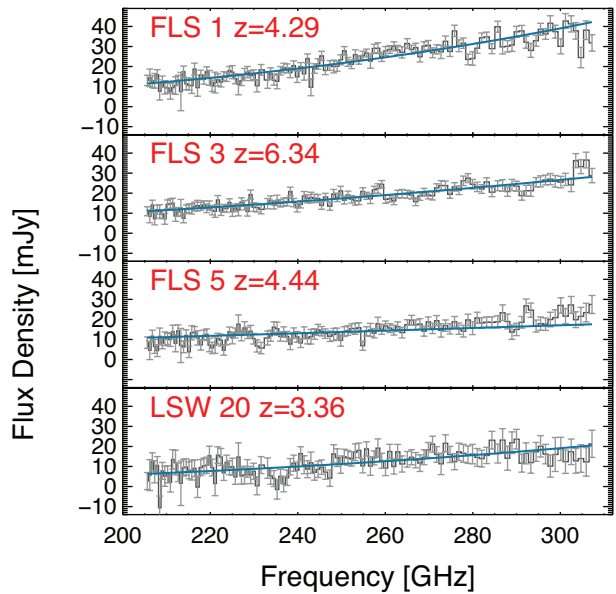}
\caption{Z-Spec observations of our sources.  All four have known
  redshifts from either CARMA or Keck as described in the
  text. Line searches in these data will be discussed in 
  Riechers et al.\ 2013, in preparation.\label{fig:zspec} }
\epsscale{1}
\end{figure}

\nobreak 
Four of the sources in our final catalog were observed with Z-Spec on
the Caltech Submillimeter Observatory (CSO) in 2010 March-May during
favorable conditions ($\tau_{\mathrm{225\,GHz}}$(zenith) = 0.05 -- 0.11): FLS~1,
FLS~3, FLS~5, and LSW~20.  Z-Spec is an $R \approx 250$ grating
spectrometer covering the full spectral range from $\lambda = $0.97 to
1.58\,mm \citep{ear06}, with a beam FWHM ranging from 25 to
33\arcsec\ over the band.  All of the sources were detected, with
1.1\,mm continuum flux densities of 10--30 mJy.  The spectra for FLS~1,
FLS~3, FLS~5, and LSW~20 are shown in Figure~\ref{fig:zspec} and
Table~\ref{tbl:mmflux}.  We performed a search for lines in these
spectra using the redshift search algorithm detailed in
\citet{Lupu:2012}, but did not detect any features at greater than
$3\sigma$.  These spectra are analyzed further by Riechers et al.\ 2013 (in
preparation).

Two of our targets, LSW~48 and LSW~52, were observed with Bolocam at
1.1\,mm in December 2012 under excellent conditions
($\tau_{225GHz}$(zenith) = 0.03 -- 0.06) using a Lissajous scan
pattern.  Bolocam is a facility 144-element bolometer camera at the
CSO that can operate at either 1.1 or 2.1 mm \citep{Glenn:03}.  The
beam FWHM at 1.1 mm is 31\arcsec , and the fractional bandwidth is
0.17.  The total integration time for LSW~48 was 2.0 hrs, and for
LSW~52 was 4.3 hrs.  The reduction procedures are as described in
\citet{wu:12}.  Both sources were detected, and are unresolved by the
Bolocam beam.

\subsection{Sub-mm/mm Interferometric Continuum Observations: SMA and PdBI}
\label{subsec:interfer}
\nobreak 
Four sources (FLS~1, FLS~3, FLS~5, and LSW~20) were observed with the
Submillimeter Array (SMA) at $\lambda$ = 1.1\,mm using a combination of
compact, sub-compact, and extended configurations.  Similarly, nine of
our targets (FLS~1, FLS~3, FLS~6, FLS~17, FLS~19, LSW~20, LSW~28,
LSW~29, and LSW~102) were observed with the Plateau~de~Bure
Interferometer (PdBI) at $\sim 1.3$ mm.  In the majority of the cases
-- including all five with redshifts -- the large millimeter-wave
fluxes are confirmed, and in most cases the emission at
3\arcsec\ resolution is dominated by a single spatial component.
Further details of these observations, including source morphologies
and more precise interferometric positions, will be presented by
Clements et al.\ (in preparation) and Perez-Fournon et al.\ (in
preparation), respectively.

The SMA and PdBI detections allow us to measure the positional
accuracy of our SPIRE detections.  For those sources -- which tend to
be the brighter members in the sample -- the median separation of the
interferometric source and the centroid of the source in the SPIRE
difference map is 5.1\arcsec .  For sources which are isolated and
clearly detected in the 250 or 350\,\mum\ bands, a significantly
better SPIRE position can be derived by using those images alone.  For
fainter and blended sources, the positional accuracy is probably
somewhat worse.

The SMA and PdBI flux densities of 8 of our sources were measured from
Gaussian fits in the {\it uv}-plane, and are given in
Table~\ref{tbl:mmflux}; final flux densities for the other sources are
awaiting the completion of all scheduled observations.  LSW~20
is resolved into two sources with a separation of $\sim 3.8\arcsec$ by
the SMA observations (also see \S\ref{subsec:keck}).  Because the
SPIRE and Z-Spec beams are much larger than this ($>18\arcsec$), we
add the flux densities of the two components for our analysis.  The
PdBI observations of this source also show indications of a faint
component at the same position, but the flux density is poorly
constrained.  Combined with the Z-Spec and Bolocam observations, we
can see that, with the exception of FLS 17, our targets are among the
strongest known optically-faint dusty galaxies at $\lambda \approx$ 1
mm that are not known to be significantly lensed.

\subsection{Radio Continuum Observations}
\nobreak 
\label{subsec:radio}
FLS~3 was the subject of deep targeted Jansky Very Large Array (JVLA)
and follow-up at 21cm once its redshift was determined \citep{rie13}.
A radio source is detected at $6\sigma$ in the deep \citet{mor10} JVLA
survey of GOODS-N close to the position of GOODSN~8.  There are no
other detected sources within 15\arcsec.  However, without a precise
sub-mm/mm interferometric position for this source (unlike FLS~3), we
are unable to determine if this is, in fact, a radio detection of our
source.  LSW~26, which lies within the region covered by the deep
\citet{owen08} 20\,cm survey of the Lockman-Hole, is not detected in
those data, but without knowing the source size the exact detection
limit is uncertain.  

Enhanced Multi-Element Radio Linked Interferometer Network (e-MERLIN)
observations of four of the sources in this paper (FLS~1, FLS~3,
FLS~5, and LSW~20) were carried out in March~2012.  A central tuning
of 1.55\,GHz with an instantaneous bandwidth of 348\,MHz was used.
All seven stations were available, resulting in a
0.3\arcsec\ synthesized beam.  3C286 was used as the primary flux
calibrator and OQ208 was used as the bandpass calibrator. The sources
were observed for 6 hours each from 26th - 30th of March 2011. The
data were flagged spectrally, averaged in frequency, fringe fitted and
then calibrated using standard methods for phase referencing
experiments.  However, none of the sources were detected, with map RMS
value of $15-18\,\mu\mathrm{Jy}$.

We can compare these observations with those predicted by the
far-IR/radio correlation, which is parameterized by $q_{\mathrm{IR}}$,
the logarithmic ratio of $L_{\mathrm{IR}}$ and the 1.4 GHz flux
density.  The lack of redshifts, as well as the poor constraints on
$L_{\mathrm{IR}}$ in the absence of Rayleigh-Jeans-side
(longer-$\lambda$) data (Table~\ref{tbl-otherflux}), make this
comparison uninformative for GOODSN~8 and LSW~26.  For the other
sources, if we assume $S_{\nu} \propto \nu^{-0.8}$ in the radio, and
adopt the value for $q_{\mathrm{IR}}$ measured for $z\sim 2$ {\it
  Herschel}-selected DSFGs from \citet{ivison:10}, we can predict the
range of expected $S_{1.55\,\mathrm{GHz}}$ flux densities for
comparison with our detection limits.  Including the scatter in
$q_{\mathrm{IR}}$ and the measurement uncertainties in each
$L_{\mathrm{IR}}$, we predict 95\% central confidence limits of
50-324, 15-94, 21-151, and $15-116\,\mu\mathrm{Jy}$ for FLS~1, FLS~3,
FLS~5, and LSW~20, respectively.  The non-detections of FLS~3, FLS~5
and LSW~20 are consistent with the far-IR/radio correlation, while
FLS~1 is fainter in the radio than expected at moderate significance.
However, the e-MERLIN non-detection of FLS~3 also disagrees with the
JVLA $1.4\,\mathrm{GHz}$ measurement ($S_{1.4\,\mathrm{GHz}} = (59 \pm
11) \mu$Jy), which is consistent with the far-IR/radio correlation
\citep{rie13}.  This discrepancy suggests that e-MERLIN may be
partially resolving out the radio emission for these sources.

\subsection{Optical Spectroscopy: Keck and GTC}
\label{subsec:keck}
\begin{figure*}[t!]
\centering
\includegraphics[width=0.8\textwidth]{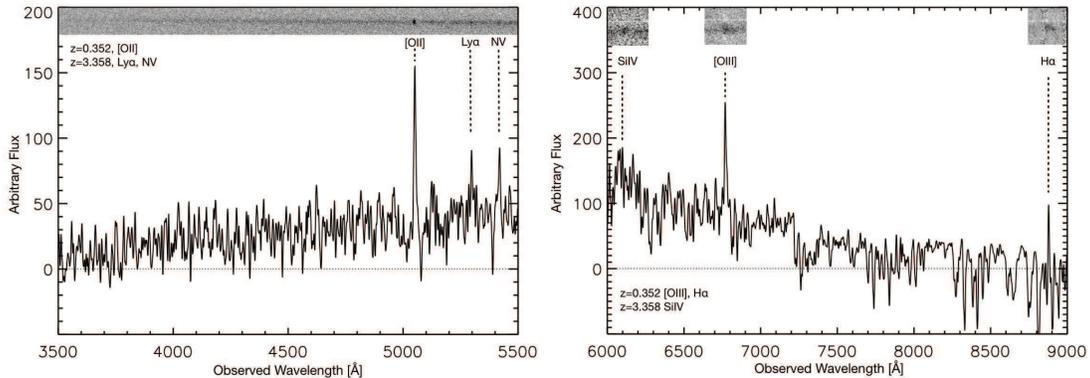}
\caption{The blue (left) and red (right) Keck/LRIS spectra of LSW
  20. Two sources are detected in multiple lines -- a $z=0.352$
  foreground source, and a background $z=3.358$ source.  The CO
  redshift of the DSFG is $z=3.32$.\label{fig-locksw20} }
\end{figure*}

\nobreak 
In addition to the sub-mm through radio observations, FLS~1 and LSW~20 were
targeted for optical spectroscopy in July 2010 and May 2011 using the
Low Resolution Imaging Spectrograph \citep[LRIS;][]{Oke:95} on the
Keck~I 10\,m telescope, making use of their high-precision SMA
interferometric positions.  We utilized a 1\arcsec\ slit width, and
the D560 dichroic with the 4000\AA, 600\,line\,mm$^{-1}$ grism (blue side)
and 400\,line\,mm$^{-1}$ grating (red side) blazed at
$\sim$7800\AA\,. The data were binned ($1\times2$) in the spectral
direction on the blue side and unbinned on the red, providing similar
wavelength dispersions of 1.26 and 1.16\AA\, per pixel, respectively.
The targets were observed with median seeing of $\sim$1\farcs0, with a
total integration time of 2.2 and 1.2 hrs, respectively.  The LRIS
data were reduced using standard {\tt IRAF} procedures.  No continuum
or emission lines were detected toward the SMA position for FLS~1.
However, spectroscopy of LSW~20 (Figure~\ref{fig-locksw20}) reveals
multiple emission lines from both a foreground $z=0.352$ and high
redshift $z=3.358$ galaxy, both toward the brighter component in the
SMA image.  No continuum or emission lines were detected from the
fainter SMA component.  This possible lensing system will be discussed
further elsewhere.  FLS~3 was also observed with LRIS, as discussed
in \citet{rie13}.

Optical spectroscopic observations of FLS~1 were also carried out on
16 May 2013 (UT) using the Optical System for Imaging and low
Resolution Integrated Spectroscopy
(OSIRIS)\footnote{http://www.gtc.iac.es/instruments/osiris/osiris.php}
on the Gran Telescopio Canarias (GTC) 10.4m telescope.  Observing
conditions were photometric, with a median seeing of 0.75\arcsec, and
the total integration time was 1.5 hrs. We utilized a 1.2\arcsec\ slit
width, and the high throughput R2500R VPH grism, which provides a
spectral range of 5630 - 7540\,\AA\ and a wavelength dispersion of
1.04 \AA\ per pixel. The slit was positioned at an angle of 343.5 deg
including a faint optical object close to the radio interferometric
position and a faint galaxy at 1 arcmin to the NW.  No continuum or
emission lines were detected close to the FLS~1 position. However,
faint continuum and emission lines were detected at RA 17h08m15.727s,
Dec +58d29m29.35s, about 52\arcsec\ from FLS~1 and 8\arcsec\ from the
galaxy used to position the slit.  Careful analysis of the 2D spectrum
at this position reveals emission lines (H$\beta$ and [\ion{O}{3}]
4959, 5007\,\AA\ ) from a low redshift galaxy at $z=0.415$ and a
narrow line at 6430\,\AA\ which we identify as Lyman$\alpha$ at
$z=4.287$ (Figure~\ref{fig-fls1gtc}), very close to the CO redshift of
FLS~1 ($z=4.286$; see below).  The serendipitous detection of a
galaxy at the same redshift as FLS~1 in a narrow long-slit is highly
improbable unless there is an overdensity of galaxies close to FLS~1
as detected around other high redshift SMGs (e.g. Capak et al. 2011).
The GTC observations will be presented in more detail elsewhere.

\begin{figure}
\epsscale{1.1}
\plotone{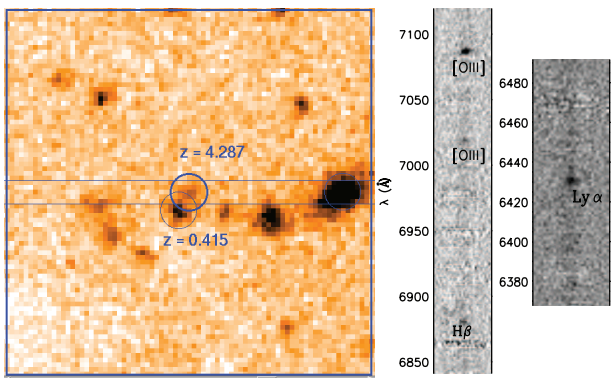}
\caption{ GTC observations of galaxies near FLS~1.  The left panel
  shows a GTC $z$-band image of the field around a high redshift
  galaxy discovered during the long-slit observations of FLS~1. The
  spectra of two galaxies overlap in the long-slit spectrum.  The slit
  is shown by the horizontal lines, and the circled galaxy on the
  right was used to align the slit.  The middle panel shows the 2D GTC
  OSIRIS spectrum of the $z = 0.415$ galaxy showing emission lines of
  H$\beta$ and [\ion{O}{3}] 4959, 5007\,\AA , and the right panel the
  spectrum of the $z=4.287$ galaxy showing narrow Ly$\alpha$ emission.
  FLS~1 is 30\arcsec\ outside the frame of the $z$-band image.
\label{fig-fls1gtc}
}
\epsscale{1}
\end{figure}

\subsection{CARMA Millimeter-Wavelength Line Search}
\nobreak 
Five of our sources have been observed with CARMA with the main goal
of redshift detection.  These observations are described by Riechers
et al.\ 2013 (in preparation).  The redshift for LSW~20 has been
confirmed with a CO line; the measured CO value is $z = 3.32$,
corresponding to a relative velocity of 2600 km/s in the source frame
with respect to the optical redshift.  The source of this discrepancy
is currently not clear.  FLS~1 has a redshift of 4.29 based on two CO
lines.  FLS~3 has a redshift of 6.34 based on multiple emission lines
\citep{rie13}.  FLS~5 has a redshift of 4.44 based on multiple CO
lines, and LSW~102 has a redshift of 5.29 based on clear detections of CO
J(6-5), (5-4), and [\ion{N}{2}] 205\,\mum .  The uncertainties in
these redshifts are negligible for the purposes of this paper.

\begin{table*}[t!]
 \centering
 \caption{Additional follow-up}
 \begin{tabular}{l|c|rrrrr|r|r|r|c}
 \hline
  Source & Optical & \multicolumn{5}{c}{Z-Spec} &
  Bolocam & SMA & PdBI & Radio \\
   & & $S_{1016\,\mum}$ & $S_{1113\,\mum}$ & $S_{1210\,\mum}$ & $S_{1311\,\mum}$ & 
   $S_{1411\,\mum}$ & $S_{1.1\,mm}$ & $S_{1.1\,mm}$ & $S_{1.3\,mm}$ & \\
\hline\hline
FLS 1  & \checkmark & $34.1 \pm 0.7$ & $28.3 \pm 0.7$ & $21.8 \pm 0.5$ & $15.0 \pm 0.6$ & $12.1 \pm 0.8$ & \nodata & $13.8 \pm 0.8$ & $14.5 \pm 0.9$ & \checkmark \\
FLS 5  & & $20.3 \pm 0.7$ & $16.6 \pm 0.5$ & $12.5 \pm 0.6$ & $11.4 \pm 0.6$ &  $9.5 \pm 0.7$ & \nodata & $17.2 \pm 1.1$ & \nodata & \checkmark \\
LSW 20 & \checkmark & $16.8 \pm 1.1$ & $14.9 \pm 0.8$ & $12.0 \pm 0.8$ & $7.0 \pm 0.9$  & $7.2 \pm 1.1$  & \nodata & $18.4 \pm 1.3$ & $8.7 \pm 1.8$ & \checkmark \\
\hline
FLS 6  & & \nodata & \nodata & \nodata & \nodata & \nodata & \nodata & \nodata & $8.4 \pm 0.6$ & \\
FLS 17 & & \nodata & \nodata & \nodata & \nodata & \nodata & \nodata & \nodata & $1.5 \pm 0.4$ & \\
FLS 19 & & \nodata & \nodata & \nodata & \nodata & \nodata & \nodata & \nodata & $9.1 \pm 0.8$ & \\
LSW 29 & & \nodata & \nodata & \nodata & \nodata & \nodata & \nodata & \nodata & $8.4 \pm 1.9$ & \\
LSW 48 & & \nodata & \nodata & \nodata & \nodata & \nodata & $13.0 \pm 2.4$ & \nodata & \nodata & \\
LSW 52 & & \nodata & \nodata & \nodata & \nodata & \nodata & $7.5 \pm 1.6$ & \nodata & \nodata & \\
\hline
\end{tabular}
\tablecomments{ Additional observations for our sources. The optical
  spectroscopy and radio observations are discussed in
  \S\ref{subsec:keck} and \S\ref{subsec:radio}, respectively.  The
  mm-band flux densities are given in mJy, and only instrumental noise
  is included in the quoted uncertainties.  Observations of FLS 3 are
  described in \citet{rie13}.  Sources in the top portion of the table
  have known redshifts.  The SMA flux density for LSW 20 is the sum of
  the two components, but for PdBI is only for the brighter of the two
  components.  The Z-Spec flux densities are binned into five equal
  sized bins in wavelength. \label{tbl:mmflux} }
  \vskip 0.1cm
\end{table*}


\subsection{Fits to SPIRE plus Sub-mm/mm Photometry}
\nobreak 
\label{subsec:detailed-fits}
Here we present modified blackbody fits to FLS~1, FLS~5, and LSW~20,
all of which have known redshifts and mm-band observations to
constrain the red side of the thermal SED.  We use the same MCMC code
as described in \S\ref{sec:basic} and the Appendix, but omit the
priors on $\beta$ and $\lambda_0$.  A similar analysis for FLS~3 is
presented by \citet{rie13}.  The mm-band observations of LSW~102 are
ongoing. We use the results to constrain the dust temperature, far-IR
luminosity (and hence star-formation rate), and dust mass for these
sources.  The calibration uncertainties and confusion noise are
handled as previously for SPIRE data.  In addition, we assume 10\%
calibration uncertainties for SMA, PdBI, and Z-Spec that are
uncorrelated between observatories, but are perfectly correlated
between the Z-Spec channels.  Given the high resolution of the
interferometric observations, we do not include confusion noise for
the SMA or PdBI observations, since it is negligible compared with
the other uncertainties.  We expand our estimates of the
correlation in the confusion noise to the binned Z-Spec channels using
the same simulations as for the SPIRE bands.  The Z-Spec and SMA
1.1\,mm flux densities for FLS~1 disagree significantly.  We have been
unable to determine why this is the case, but given the good agreement
for this source between the Z-Spec and PdBI 1.3\,mm observations, we
do not use the SMA observations for this source.  As
noted in \S\ref{sec:basic}, the blue-side power law
($\lambda^\alpha$), does not improve our fits.

In addition to the SED parameters, we measure $L_{\mathrm{IR}}$ (the
integrated IR luminosity between 8 and 1000\,\mum\ in the rest frame),
and the dust mass, $M_{\mathrm{d}}$.  For these calculations, and
elsewhere in this paper, we assume a flat Universe and a cosmological
constant with $\Omega_{\mathrm{m}}=0.27$ and $H_0 =
70\,\mathrm{km}\,\mathrm{s}^{-1}\,\mathrm{Mpc}^{-1}$.  The dust mass
is estimated using
\begin{displaymath}
 M_{\mathrm{d}} = S_{\nu} D_{\mathrm{L}}^2 \left[\left(1+z\right)
  \kappa_{\nu} B_{\nu}\right]^{-1} \tau_{\nu} \left[1-\exp\left(-\tau_{\nu}\right)
\right]^{-1},
\end{displaymath}
where $S_{\nu}$ is the flux density, $D_{\mathrm{L}}$ the luminosity
distance, $\kappa_{\nu}$ the mass absorption coefficient, $\tau_{\nu}$
the optical depth, and $B_{\nu}$ the Planck function, with all
quantities expressed in the observer frame.  For the mass absorption
coefficient we adopt $\kappa_{\nu} =
2.64\,\mathrm{m}^2\,\mathrm{kg}^{-1}$ at 125\,\mum\ \citep[rest
  frame,][]{dunne03}.  Our quoted uncertainties on $M_{\mathrm{d}}$ do
not include the uncertainty in $\kappa_{\nu}$, which is at least a
factor of 0.3\,dex.  To compute the star formation rate, we use the
relation of \citet{kenn98}, which assumes a Salpeter IMF.  Note that
$L_{\mathrm{IR}}$, $M_{\mathrm{d}}$, and the star formation rate
assume no lensing magnification.  Lensing constraints from
interferometric observations will be described in future papers.

\begin{figure*}[t!]
 \centering
 \includegraphics[width=0.8\textwidth]{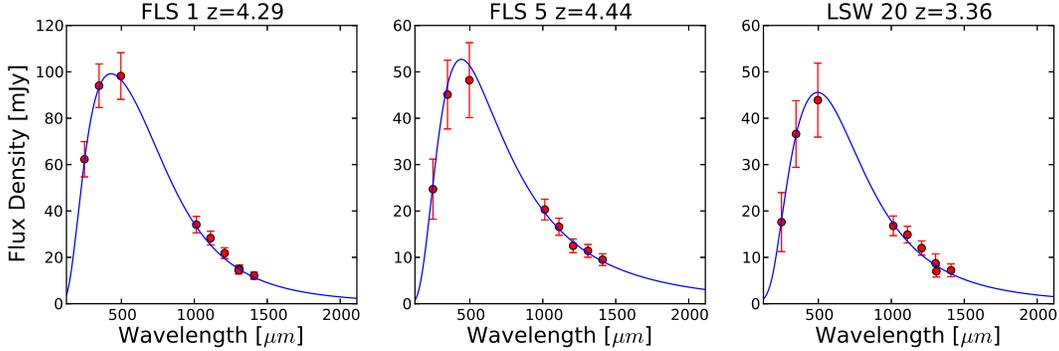}
\caption[Detailed fits]{Modified blackbody dust SED fits to sources with
  known redshifts, as described in \S\ref{subsec:detailed-fits}.  The
  displayed uncertainties include calibration and confusion noise.
  \label{fig:detailedfits}}
\end{figure*}

The results of these fits are given in Table~\ref{tbl:detailed} and
shown in Figure~\ref{fig:detailedfits}.  The derived properties of
these sources ($T_d$, $\beta$) are fairly typical of the most luminous
{\it Herschel}-selected DSFGs at lower $z$.  The $\beta$ values are
somewhat higher than usual, but consistent, given the uncertainties,
with the often assumed value $\beta = 2$. Due to the large
uncertainties in $\beta$ for LSW~20, $M_{\mathrm{d}}$ and the area of emission
($A_{\mathrm{em}}$) are not well constrained, so only lower limits on these two
quantities are provided.  The $\chi^2$ values are acceptable for all
fits.  The fits favor using an optically thick model, with a typical
decrease in the $\chi^2$ of about 3 for adding one additional parameter.
However, the data are not sufficient to rule out an optically thin
model for any of these sources.

The derived $L_{\mathrm{IR}}$ values suggest that these sources are
among the most luminous IR sources known (assuming no magnification).
Unlike FLS~3 \citep{rie13}, where the data is consistent with an
optically thin model, the SEDs for all three sources presented here
moderately favor becoming optically thick near
200\,\mum\ (rest-frame).  Therefore, CO excitation modeling of these
sources may have to consider extinction.

\begin{table*}[t!]
 \centering
 \caption{Fits to sources with known redshifts.}
 \begin{tabular}{llll}
 \hline
 Parameter & Value & Description & Note \\
\hline\hline
\multicolumn{3}{c}{FLS 1, $z=4.29$} & \\
\hline
  $\chi^2$ & 4.83 & $\chi^2$ & 
   5 DOF; $P\left(>\chi^2\right) = 0.44$ \\
  $T_{\mathrm{d}}$ & $63^{+3}_{-4}$ K & Dust temperature & \\
  $\beta$ & $2.8 \pm 0.6$ & Extinction slope & \\
  $f_{500\,\mum}$ & $91 \pm 11$ mJy & Normalization at 500\,\mum &
   Observer frame \\
  $\lambda_0$ & $200^{+26}_{-23}$\,\mum &
   $\lambda$ where $\tau = 1$ & Rest frame \\
  $\alpha$ & $>3.6$ & Blue side power-law slope &
   68.3\% limit \\
  $L_{\mathrm{IR}}$ & $5.6^{+0.7}_{-0.9} \times 10^{13}$ $\mathrm{L}_{\odot}$ & 
  IR luminosity & 8--1000\,\mum\ (rest frame) \\
  $M_{\mathrm{d}}$ & $5.1^{+1.6}_{-2.4} \times 10^9$ $\mathrm{M}_{\odot}$ &
  Dust mass & \\
  SFR & $9700^{+1200}_{-1600}$ $\mathrm{M}_{\odot}\,\mathrm{yr}^{-1}$ &
  Star formation rate & \\
  $A_{\mathrm{em}}$ & $82^{+12}_{-21}$ $\mathrm{kpc}^2$ & 
  Dust emission area & \\
\multicolumn{3}{c}{FLS 5, $z=4.44$} & \\
\hline
  $\chi^2$ & 2.5 & $\chi^2$ 
  & 4 DOF; $P\left(>\chi^2\right) = 0.64$ \\
  $T_{\mathrm{d}}$ & $59 \pm 6$ K & Dust temperature & \\
  $\beta$ & $1.7 \pm 0.5$ & Extinction slope & \\
  $f_{500\,\mum}$ & $47 \pm 7$ mJy & Normalization at 500\,\mum &
   Observer frame \\
  $\lambda_0$ & $191 \pm 62$\,\mum &
   $\lambda$ where $\tau = 1$ & Rest frame \\
  $\alpha$ & $>3.9$ & Blue side power-law slope &
   68.3\% limit \\
  $L_{\mathrm{IR}}$ & $2.8^{+0.5}_{-0.6} \times 10^{13}$ $\mathrm{L}_{\odot}$ & 
  IR luminosity & 8--1000\,\mum\ (rest frame) \\
  $M_{\mathrm{d}}$ & $2.0^{+0.6}_{-0.8} \times 10^9$ $\mathrm{M}_{\odot}$ &
  Dust mass & \\
  SFR & $4800 \pm 900$ $\mathrm{M}_{\odot}\,\mathrm{yr}^{-1}$ &
  Star formation rate & \\
  $A_{\mathrm{em}}$ & $67^{+26}_{-31}$ $\mathrm{kpc}^2$ &  
  Dust emission area & \\
\multicolumn{3}{c}{LSW 20, $z=3.36$} & \\
\hline
  $\chi^2$ & 6.0 & $\chi^2$ 
  & 5 DOF; $P\left(>\chi^2\right) = 0.31$ \\
  $T_{\mathrm{d}}$ & $48 \pm 6$ K & Dust temperature & \\
  $\beta$ & $2.8 \pm 1.2$ & Extinction slope & \\
  $f_{500\,\mum}$ & $41 \pm 8$ mJy & Normalization at 500\,\mum &
   Observer frame \\
  $\lambda_0$ & $210 \pm 59$\,\mum &
   $\lambda$ where $\tau = 1$ & Rest frame \\
  $\alpha$ & $>3.9$ & Blue side power-law slope &
   68.3\% limit \\
  $L_{\mathrm{IR}}$ & $1.1^{+0.2}_{-0.3} \times 10^{13}$ $\mathrm{L}_{\odot}$ & 
  IR luminosity & 8--1000\,\mum\ (rest frame) \\
  $M_{\mathrm{d}}$ & $>2.0 \times 10^9$ (95\%) &
  Dust mass & \\
  SFR & $1900^{+300}_{-500}$ $\mathrm{M}_{\odot}\,\mathrm{yr}^{-1}$ &
  Star formation rate & \\
  $A_{\mathrm{em}}$ & $>32$ $\mathrm{kpc}^2$ (95\%) & Dust emission area & \\
\hline
\end{tabular}
\tablecomments{ The results of modified blackbody fits to FLS~1, 
FLS~5, and LSW~20, as detailed in the text. $L_{\mathrm{IR}}$, 
$M_{\mathrm{d}}$, $A_{\mathrm{em}}$,
and the SFR assume no lensing magnification. \label{tbl:detailed}}
\end{table*}


\section{The Number Density of Red SPIRE Sources}
\label{sec:numdens}
\nobreak 
In order to estimate the space density of bright, red SPIRE
sources, we must determine both how efficient our selection is at finding
red sources and the expected false detection rate.  Given the modest
number of sources detected, and the large observational uncertainties
in the SPIRE colors, we do not attempt to measure the differential
number counts as a function of color or flux density, but simply the
number of sources per deg$^2$.  If we denote the efficiency for the
detection of the i$^{\mathrm{th}}$ source by $\epsilon_i$ and the expected
purity of our catalog by $p$ (so $1-p$ is the fraction of the detected
sources that do not belong in our sample), then the space density of
sources $N$ is
\begin{equation}
 N = \frac{p}{A} \, \sum_i \frac{1}{\epsilon_i},
\label{eqn-nsrc}
\end{equation}
where $A$ is the area of the survey and the sum is over the sources in
our final catalog.  We now discuss how we compute $\epsilon_i$ and $p$.

\subsection{Efficiency}
\label{subsec:efficiency}
\nobreak 
Our approach for computing the efficiencies is to work as closely as
possible with the observables rather than try to compress the
information by assuming some SED model and reducing the problem to a
smaller number of variables (e.g., assuming an optically thin, fixed
$\beta$ modified blackbody and computing $\epsilon$ as a function of
$T_d$ and a suitable normalization variable).  It is not clear how
well the data are described by such simple models, particularly on the
blue side of the thermal SED \citep[see, e.g.,][]{blain03}.
Furthermore, even if they were perfectly described in the mean by such
a simple model, the degree of intrinsic deviation of individual DSFGs
from such models is not well characterized.

Therefore, we measure the efficiency directly using the observed flux
density of each source in the SPIRE bands, which should be
significantly more accurate than using an SED model.  Because our
efficiency is a complicated mix of these variables, however, we can
not make any statement that we are complete up to some simply
expressible threshold.  Thus, while our model should be substantially
better than one based on a simple SED model, it is not trivial to
visualize how $\epsilon$ depends on the SPIRE flux densities.

\begin{figure}
\plotone{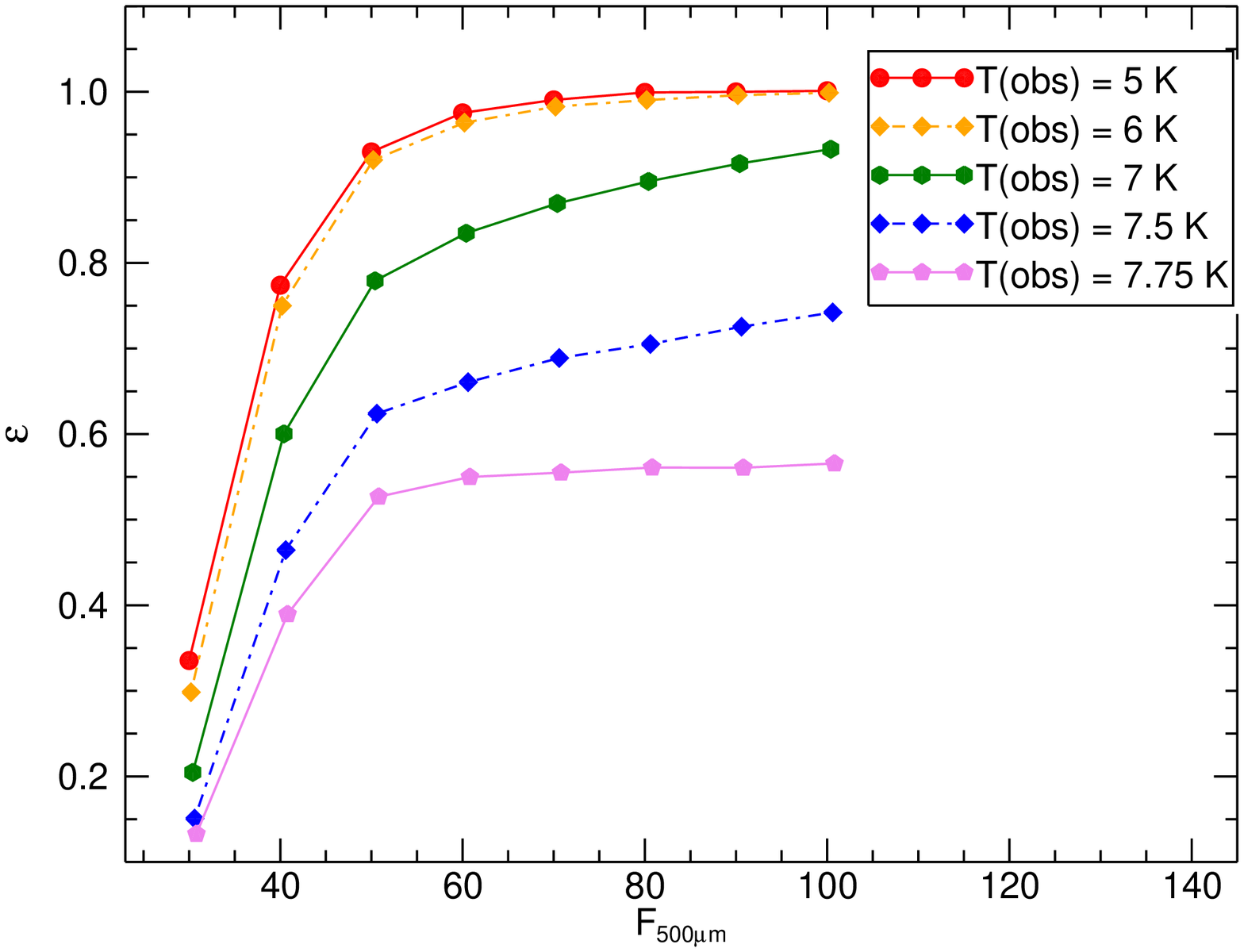}
\caption[Selection Efficiency Illustration]{Illustration of our
  selection efficiency for a simple optically thin modified blackbody
  (with $\beta=1.5$) plus \mbox{blue-side} power law model ($S_{\nu}
  \propto \nu^{-4}$) as a function of $\fplw$ and the observer frame
  dust temperature $T\left(\mathrm{obs}\right) = T_d /
  \left(1+z\right)$.  The color ratios are a strong function of
  temperature, and, for these values of $\beta$ and $\alpha$, sources
  with $T\left(\mathrm{obs}\right) > 7.75$ K are not ``red''.  For
  sources with $\fpmw \simeq \fplw$ the selection efficiency is
  relatively independent of flux density, since the dominant factor is
  the uncertainty in whether the source actually satisfies $\fplw \geq
  \fpmw$ due to instrument and confusion noise.  These simulations
  were carried out for the FLS field. \label{fig:teff}}
\end{figure}

Therefore, purely for this purpose, we illustrate our
selection efficiency for a modified blackbody SED model as a function
of temperature and \fplw\ in
Figure~\ref{fig:teff}, using the approach detailed below.  The model
is an optically thin modified blackbody coupled with a $nu^{-\alpha}$ power
law (on the Wien side, fixing $\beta = 1.5$ and
$\alpha = 4$ based on the properties of nearby (hence well observed)
IR-luminous galaxies.  Note that many of our actual sources are
not well fit by these parameters, and so we do not use this
model in any way while computing the actual detection efficiencies.
This figure does, however, correctly illustrate that the detection
efficiency decreases quite rapidly at low flux densities, which is the
basis of our selection requirement $\fplw \ge 30$ mJy.

Due to the relatively small number of sources in our catalog, rather
than attempt to compute the efficiencies on a grid of the three flux
densities (for each field) and then interpolate, it is more efficient
to compute them individually for each source.  Our basic tool for
doing so is, for each source in our catalog, to inject false sources
into the actual maps for each field with properties that match the
measurements of each source, and then determine how many are recovered
using the search pipeline.

An important point is that we do not know the precise flux densities
of any source due to observational uncertainties (instrument and
confusion noise), and therefore each efficiency has an associated
uncertainty which can not be reduced without additional observations.
We include this effect in our simulations by generating a moderate
number ($N_{\sigma}$) of flux density triplets for each source drawn
from a multi-variate Gaussian centered on the observed flux densities
and using the estimated uncertainties.  We then carry out
$N_{\mathrm{sim}}$ simulations for each of the flux density triplets,
each of which has $N_{\mathrm{src}}$ sources injected, averaging to form
$N_{\sigma}$ estimates of $\epsilon$ for each source.  The final
$\epsilon$ is then the average of the $N_{\sigma}$ values, and the
uncertainty is given by the distribution of values.

$N_{\sigma}$ does not have to be very large to provide a good estimate
of the uncertainty, and so we generally adopt a value of $N_{\sigma} =
60$, except for a few sources where we used $N_{\sigma} > 1500$ in
order to verify convergence.  $N_{\mathrm{src}}$ is limited by the
desire to avoid two sources overlapping within the matching radius,
$m_{\mathrm{rad}}$.  We have conservatively adopted $m_{\mathrm{rad}}
= 15\arcsec$, compared with the typical positional uncertainty of
5\arcsec\ for our faintest sources in simulations (and from
interferometric observations: \S\ref{subsec:interfer}).  With these
values, the correction for sources lost due to positional
uncertainties are negligible compared with our other uncertainties.
We adopt $N_{\mathrm{src}} = 800$ for our FLS simulations,
$N_{\mathrm{src}} = 2500$ for those of the combined Lockman fields, and
$N_{\mathrm{src}} = 50$ for GOODS-N, which limits the effects of
overlap on the recovered counts to less than 1\%, again much smaller
than our other sources of uncertainty.  It is important that
$N_{\mathrm{sim}} \times N_{\mathrm{src}}$ is large enough that
$\epsilon$ is measured for each flux triplet to much better precision
than the variation between flux triplets for the same source.  We have
adopted $N_{\mathrm{sim}} = 60, 30, 800$ for FLS, Lockman, and
GOODS-N, respectively, based on tests with much larger
$N_{\mathrm{sim}}$ values.  In total, therefore, the efficiency for
each FLS source is based on 2.4 million injected sources, GOODS-N on
2.4 million, while in Lockman that value is 4.5 million; more
sources are desirable for the Lockman fields because the noise
properties of the map are less uniform.

\begin{figure}
\plotone{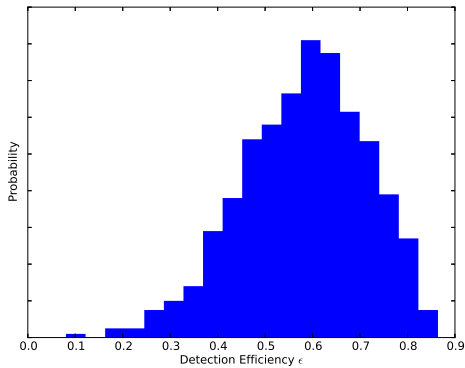}
\caption[Histogram of efficiencies]{Histogram of efficiencies
  ($\epsilon$) for FLS 5.  Each entry represents a different flux
  density triplet reflecting the observational (instrumental and
  confusion) uncertainties.  For this source, $N_{\sigma} = 1500$
  simulations were carried out.  The resulting efficiency has significant
  non-Gaussian tails. \label{fig:effhist}}
\end{figure}

The resulting efficiencies and uncertainties are provided in
Table~\ref{tbl-otherflux}.  They range from $\sim 90\%$ for FLS~3 (the
$z=6.34$ source) to $\sim 20\%$ for faint sources with $\fplw \simeq
\fpmw \simeq \fpsw$.  The uncertainties in each $\epsilon_i$ are
generally non-Gaussian (Figure~\ref{fig:effhist}), especially for
sources with $\fplw \simeq \fpmw$.  We use the measured $\epsilon$
distribution for each source when computing the source density, as
discussed below.  Again, note that the uncertainties in the $\epsilon$s
primarily arise from the uncertainties in the flux densities of each
source, not from the finite number of simulations we carried out, and
hence are irreducible given current observations.

\subsection{Purity}
\label{subsec:purity}
\nobreak 
There are two classes of contaminants in our analysis: faint red
sources below our $\fplw \ge 30$ mJy cut, and non-red sources.  The
effects of the former are commonly referred to as `boosting' or
`Eddington bias' in the sub-mm/mm literature, and can be caused by
both instrument noise and blending effects, while those of the latter 
almost entirely arise from instrument noise because blending tends to
make sources bluer.  The rate of contamination of fainter
sources is easy to estimate using the same methods as in
\S\ref{subsec:efficiency}, but the second is much more difficult to
characterize.  False detections caused by map artifacts, evaluated
through jack-knife tests on our maps, are negligible.

\subsubsection{Boosting}
\label{subsubsec:boosting}
\nobreak 
There are two ingredients required to estimate the contamination by
red but faint ($\fplw < 30$ mJy) sources: a measurement of the
efficiency of detecting such sources; and an estimate of their number
density.  For the former, we use the same technique as in the previous
section to measure the detection efficiency as a function of \fplw ,
omitting the $N_{\sigma}$ sampling over photometric errors.  Here we
must assume SPIRE color ratios.  We adopt the median of our catalog
sources, $\fplw / \fpmw = 1.09$ and $\fplw / \fpsw = 1.88$, and later
check to see how changing these values affects our results.  The
detection efficiency would be essentially zero in the absence of
boosting effects, and is a strong function of \fplw .  We perform a
series of simulations, sampling $\fplw$ from 29 to 2 mJy in 1 mJy
steps.  As expected, the recovery efficiency falls quite rapidly, from
$8.9\%$ at $\fplw=29$ mJy to $<0.01\%$ at 5 mJy, crossing 1\% at 18
mJy.  We carry out separate simulations for each field with at least
250,000 sources per mJy bin per field, and find that the values are
essentially identical for all fields.

Unfortunately, we do not yet have good measurements of the number
counts of red sources, especially at faint flux densities.  Our
approach is therefore to scale from the observed SPIRE number counts,
which are expected to be dominated by sources at lower $z$.  Assuming
our source population has $\left<z\right> = 4.7$ (\S\ref{subsec:pz}),
a flux cut of $\fplw \geq 30$ mJy corresponds to a cut of $\fpmw \geq
61$ mJy at $z=3$, or $\fpsw \geq 140$ mJy at $z=1.9$.  The DSFG
population is thought to peak closer to $z=3$ \citep{chapman05,
  weiss:13}, so we adopt the 350\,\mum\ counts for our calculations, using the
250\,\mum\ value as a comparison to estimate the uncertainty.  For
these calculations we adopt the differential number counts in each
band from \citet{oliver10}, normalizing them to match the observed
number of red sources.

Using the above ingredients, we find an expected boosting
contamination of 0.8 sources in FLS, 1.4 in Lockman-SWIRE, and $\ll 1$
in GOODS-N, Lockman-East, and Lockman-North.  In all cases these
correspond to $\sim 0.1$ deg$^{-2}$, a small value compared to the
catalog source density of $>2$ deg$^{-2}$.  Making the simulated
sources as red as FLS 3 (our $z=6.3$ source, and one of the reddest
sources in the sample) increases this value by only 0.03 deg$^{-2}$.
Using the 250\,\mum\ counts decreases it by 0.02 deg$^{-2}$, and
using the shape of the number counts from the B11 model at $z>3$
increases it by 0.05 deg$^{-2}$.  We therefore adopt a value for the
contamination by faint red sources below our flux density cutoff of
$0.11 \pm 0.06$ deg$^{-2}$, which is negligible compared with the other
sources of uncertainty. 

\subsubsection{Non-red Contaminants}
\label{subsubsec:nonred}
\nobreak 
The basic tool for our analysis of contamination by non-red sources is
again simulations, but here rather than injecting sources into real
maps we generate simulated maps based on models of the DSFG
population, but with intrinsically red, bright sources omitted, and
run our search pipeline on those maps.  The most common type of false
detections, according to models, are {\it nearly}-red (``orange'')
sources combined with noise fluctuations or blended with faint red
sources.  Therefore, the critical model ingredient is the number of
almost-red sources.  Unfortunately, it is not clear how well current
models represent this population, given that, as discussed later, they
generally significantly under-predict the number of truly red sources.

There are a number of such models in the literature.  However, as
shown by \citet{oliver10, glenn10, Clements:10}, pre-{\it Herschel}
models do a poor job of reproducing the observed number counts at
250--500\,\mum , and therefore are not a good choice for our
simulations.  We instead make use of the post-{\it Herschel}
backward-evolution B11 model, which is a good match to current number
count constraints.  For each field, we generate a number of
simulations using this model with genuinely red sources brighter than
$\fplw = 3$ mJy removed, adding the measured noise at each position of
the map.  Simply distributing the sources randomly (i.e., in an
unclustered fashion) will underestimate the false detection rate, so
we include clustering using the measured 350\,\mum\ power spectrum of
\citet{amblard11}; note that this is purely clustering in the plane of
the sky.  If ``orange'' sources are more strongly clustered on the
sky, then this may underestimate contamination effects. However,
increasing the simulated clustering by a factor of three does not have
any measurable effect on our false detection results.  We generate 200
simulations of each field, and then run them through the same
detection pipeline as the real data, and then count the number of
sources detected.

We visually inspect all of our actual catalog sources to note blends.
We therefore must carry out the same procedure on at least a subset of
our purity simulations.  Doing so for 100 simulated maps of each field
removes approximately half the detected sources -- more precisely 83
out of 145 in FLS, and 174 of 320 in Lockman-SWIRE.  These are all
blends of sources which can clearly be identified in the un-smoothed
maps, and where the individual constituents are too faint to belong in
our catalog.  This blend rate -- 2.6 sources for FLS and Lockman-SWIRE
combined -- is consistent with the observed number (two) in our
catalog.  The false detection rates for the smaller, deeper fields
(Lockman-North, Lockman-East, and GOODS-N) are less than 0.3 per
field.

The purity is then given by $p = \left(N_{\mathrm{cat}} -
N_{\mathrm{false}}\right) / N_{\mathrm{cat}}$, where
$N_{\mathrm{cat}}$ is the number of sources in our catalog for a given
field, and $N_{\mathrm{false}}$ is the estimated number of false
detections from simulations of the same field.  Note that
$N_{\mathrm{false}}$ is determined much more precisely than
$N_{\mathrm{cat}}$, which suffers from Poisson noise.  We conclude
that the purity of our catalog sample is approximately 90\%, somewhat
higher in FLS (93\%) and somewhat lower in Lockman-SWIRE (91\%).
Interestingly, increasing the S/N requirement for detection to 5, 6,
or 7 does not have much effect on the purity -- while increasing the
S/N requirement cuts down the number of false sources in simulations,
it removes a large a fraction of real sources from the actual maps.
Some of the false detections, due to unresolved blends, are as bright
as $80$ mJy at 500\,\mum .  Decreasing the S/N requirement to 3, on
the other hand, decreases the purity to 75\%, so our value of 4 is a
good compromise.  Note that this appears to be true for all our
fields, despite their range of depths.

During the preparation of this paper but too late to be used as the
primary model for our purity simulations, the B11 model was updated
\citep[][hereafter B12]{bethermin:12}.  The B12 model predicts
significantly more ``orange'' sources than the B11 model, implying a
lower purity fraction for our catalog.  Based on 40\,deg$^2$ of
simulated data (compared with the several hundred for the B11 model),
we find a predicted number density of ``orange'' interlopers of
0.41\,deg$^{-2}$, corresponding to a purity of $\sim 80\%$.  In the
future, once more models have been updated to reflect the {\it
  Herschel} number counts, it will be helpful to compare the false
detection rates for different models to estimate the systematic
uncertainties in the purity.  The parameters of the B11
model do have estimated uncertainties, but adjusting the model
parameters to increase the number of red (and near-red) sources by
those uncertainties does not produce any detectable change in the
purity rate.  It seems likely that the systematic uncertainty term is
dominated by more fundamental assumptions related to the SED models,
etc., which we currently are not able to estimate.

\subsection{Blends} \label{subsec:blend}
\nobreak Including blends presents a problem for our source density
estimate.  In the absence of instrument noise, any blended source
which appears red must include at least one genuinely red source.
However, such a source may lie below our detection threshold, and hence
would not belong in our catalog.  Therefore, for a blend of two
sources detected at 250\,\mum, it is not clear if the source should
count as zero, one, or two sources in our density estimate.  The
probability of the first two can be estimated from models for the
color distribution of non-red sources.  Furthermore, because the
observed number density of sufficiently bright 500\,\mum-risers is
considerably smaller than non-red sources, it seems reasonable to
assume that a red source is much more likely to be blended with a blue
source than another red one -- which will make the resulting blend
bluer, but also may make it bright enough to be selected.  This
amounts to the assumption that the spatial clustering of bright red
sources is not strong enough to compensate for the much larger number of
non-red sources, an assumption that we are currently not in a position
to test.

Because available models are trained on differential number counts in
each band rather than colors (which largely have not been measured),
the uncertainty in these predictions is hard to quantify.  Therefore,
we have chosen to exclude clear blends from our source counts but
include their presence as a systematic uncertainty.  There are two
sources in our catalog that are detected as likely two-component
blends in the 250\,\mum\ data (but not in the other two bands): FLS~7
and FLS~23.  We are able to make crude estimates of the
$S_{250\,\mum}$ flux densities of the constituent sources in both
cases.  Starting with this information, we randomly select single
sources that are at least this bright at 250\,\mum\ from a large
simulation of the B11 model, but which are also fainter than the
blended source in all bands.  We then subtract the
fluxes of the simulated source from the blended fluxes of the actual
source, use this as an estimate of the flux of the other, red
component, and finally determine what fraction of the time the
de-blended source would pass our selection criteria.

For FLS~7, we find that 99.8\% of the time the de-blended source is
sufficiently bright and red to belong in our catalog.  For FLS~23,
which is significantly fainter, we find that for 110,000 simulated
companions, the resulting de-blended source belongs in our catalog
zero times.  Therefore, we conclude that FLS~7 very likely contains a
genuinely red, bright source, while FLS~23 at best contains a red source
that is too faint to pass our $D$ selection, and hence does not belong
in our catalog.

In order to estimate the systematic uncertainty due to the fact that
FLS~7 should be included in our catalog -- but with unknown true
fluxes and hence unknown efficiency -- we compute the efficiency
as above but using the ensemble of de-blended model fluxes, and then
include 99.8\% of that as a positive systematic uncertainty in our
final source density estimate.  This results in a positive systematic
uncertainty of $0.24$ deg$^{-2}$ in our FLS source density measurement.

\subsection{Source Density} 
\label{subsec:sourcedens}
\nobreak 
Combining the results of the purity and efficiency simulations with
the input catalog, we finally arrive at our measurement of the source
density of red sources in {\it Herschel} data.  One complication is
that our S/N requirement of 4 corresponds to slightly
different depths in different fields, but in simulations the S/N
seems to be a better predictor of the purity than the difference map
flux density $D$.

\begin{figure}
\epsscale{1.1}
\plotone{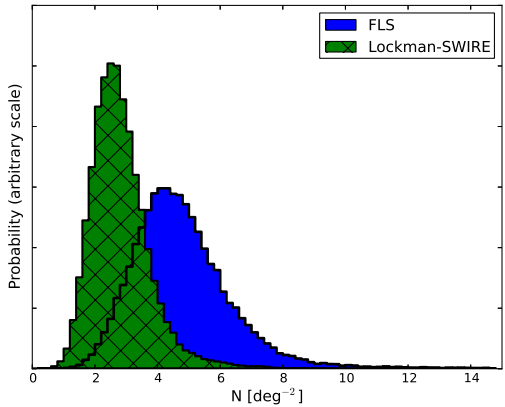}
\caption[Source density estimate for red sources from FLS and
  Lockman-SWIRE] {Estimated red source density per deg$^2$ for our two
  large fields, FLS (solid blue) and Lockman-SWIRE (hatched green),
  including all identified statistical uncertainties.  The smaller
  fields (Lockman-North, Lockman-East, and GOODS-N) are not shown
  because of their significantly different depths, and because their
  limited size (a total of 1.1 deg$^{2}$ compared with the 20.3
  deg$^{2}$ of the fields shown) means that they make very little
  contribution to our final catalogs. \label{fig:nsrc}}
\epsscale{1.0}
\end{figure}

The uncertainties in the efficiencies are quite
non-Gaussian.  To fold these into our calculation, when
computing $N$ from Equation~\ref{eqn-nsrc}, we randomly select an
efficiency for each catalog source from our $N_{\sigma}$ simulation
sets, compute $N$ using those values, and then repeat this process
several thousand times.  We also include the uncertainty in $p$ and
the Poisson uncertainties from the number of detected sources in this
step.  The result is an ensemble of values for each field, drawn from
the underlying $N$ distribution.  For the FLS field, we find that $N =
4.90^{+1.35}_{-1.47} \left(\mathrm{stat}\right)
^{+0.24}\left(\mathrm{sys}\right) \, \mathrm{deg}^{-2}$ and for
Lockman-SWIRE $N = 2.91^{+0.80}_{-0.86}\left(\mathrm{stat}\right)
\, \mathrm{deg}^{-2}$, not
including the boosting correction, and where here the systematic
uncertainties include only the effect of identified blends.  The
resulting number density distributions are shown in
Figure~\ref{fig:nsrc}.  Given these uncertainties, the two fields are in
reasonable agreement, with the difference corresponding to
$1.2\,\sigma$.  Due to the modest size of our catalogs, Poisson errors
dominate our uncertainty budget.

Since the sources in FLS and Lockman-SWIRE completely dominate our
catalog (given their much larger areas), and the difference in depth
is relatively minor considering the other uncertainties in our
computation, we combine the results from these two fields using
inverse variance weighting, then apply the boosting correction to give
our final estimate of the red-source sky density of $N =
3.27^{+0.67}_{-0.84} \left(\mathrm{stat}\right) \, \mathrm{deg}^{-2}$
for sources with $\fplw \geq 30$\,mJy and $D \geq 23.9$\,mJy, with an
identified systematic uncertainty due to partially resolved blends and
boosting of $^{+0.06}_{-0.05}\, \mathrm{deg}^{-2}$.  The corresponding
95\% lower limit is $N > 2.04$ deg$^{-2}$.  Note that these values
ignore any potential multiplicity of our sources, which will have to
be constrained with interferometric observations.  As noted in
\S\ref{subsec:interfer}, high-resolution observations of a handful of
red targets show that the multiplicity fraction appears to be
relatively low.  This is apparently at odds with the findings of
\citet{Hodge:2013} -- but this requires further study, which will be
presented by Clements et al.\ (in preparation) and Perez-Fournon et
al.\ (in preparation) using a larger sample. The number densities
should therefore be interpreted as the number density of sources at
the resolution of the SPIRE beam.

We can use the fact that 4/5 of our sources with measured redshifts
are at $z>4$ to estimate the sky density of $z>4$ DSFGs satisfying our
selection.  This assumes that the sources with redshifts are
representative of our population (see Figure~\ref{fig:clrplane}),
which may not be entirely valid.  In particular, FLS~3 and LSW~102 are
redder than most of our catalog sources, and most are brighter than
the average catalog source at 500\mum.  Evaluating the effects of this
bias requires observations of additional sources.  With this
caveat in mind, if we denote this fraction by $f_{z>4}$, and assume a
flat prior, then, from a simple application of Bayes theorem, the
probability distribution for $f_{z>4}$ is given by
$P\left(f_{z>4}\right) \propto \left(f_{z>4}\right)^4 \left(1 -
f_{z>4}\right)$.  This translates into a mean value of $f_{z>4} =
0.71^{+0.16}_{-0.17}$ (68.3\% confidence limits), with a mode of 4/5.
The corresponding frequentist confidence interval is $f_{z>4} = $
0.48--0.96.  Similarly defining the fraction of our sources with $z >
3$ ($f_{z>3}$), we can provide a 95\% Bayesian lower limit of $f_{z>3}
> 0.61$ ($>0.83$ at 68.3\%).  Combining this with the sky density
value quoted above, we therefore calculate the number density of $z>4$
red sources to be $N_{z>4}=2.37^{+0.75}_{-0.79} \, \mathrm{deg}^{-2}$,
and $N_{z>3} > 1.68 \, \mathrm{deg}^{-2}$ (95\% lower limit; we do not
provide a central confidence interval here because our constraint on
$f_{z>3}$ is a lower limit).  Note that these results do not make use of
photometric redshifts, and in fact the largest contribution to the
uncertainty budget in the number of high-$z$ DSFGs selected by our
method is due to the small number of sources with spectroscopic
redshifts.  Clearly the highest priority for future work is to obtain
additional redshifts, which will also allow detailed studies of the
physical properties of these galaxies.  None of these values attempt to
correct for non-red DSFGs at high-$z$.

\section{Other Red Sources in the Literature}
\label{sec:otherred}
\nobreak 
There are a number of 500\,\mum-riser sources in the literature with
spectroscopic redshifts, the majority of which are at $z>4$.  However,
we suspect that this sample suffers from ``publication bias'': it is
highly inhomogeneous, and the fact that these sources were singled out
for publication may have been influenced by the fact that they were at
high-$z$.  Therefore, it is very difficult to draw any quantitative
conclusions about how they relate to the population described in this
paper.  One exception is the recently published South Pole Telescope
(SPT) selected sample of bright, lensed DSFGs of \citet{vieira:13}.
Here the sources were selected at 1.4\,mm in a uniform fashion, but
SPIRE observations reveal that most of those at $z>4$ are
500\,\mum-risers.  In this section we summarize information about red
literature sources, particularly the SPT sample, which is discussed at
the end of the section.

\citet{combes12} discovered a bright ($\fplw > 200$ mJy)
500\,\mum-riser at $z=5.2$ highly magnified by the cluster Abell 773.
Similarly, \citet{cox11} discusses a strongly-lensed 500\,\mum-riser
with $z=4.24$.  Both sources would be easily detected by our method,
and in fact were selected using SPIRE data.  \citet{swin12} describes
two sources at $z = 4.4$ with redshifts serendipitously measured
during a high-resolution continuum mapping program in GOODS-S.
Examination of the deep HerMES GOODS-S map shows that both sources are
500\,\mum-risers, but are fainter than our \fplw\ cutoff.

There are also a number of known $z>4$ DSFGs detected at longer
wavelengths ($> 850$\,\mum) that are too faint to be detected in SPIRE
data (irrespective of our selection technique), such as the $z=4.8$
source described by \citet{coppin09}, or the one at $z=5.2$ from
\citet{walter:12}. Perhaps more interestingly, there are at least two
examples of $z>4$ DSFGs that are detected in the SPIRE bands, but have
$\fpmw > \fplw$: AzTEC-3 \citep[$z=5.3$][]{capak11, rie:10}, and
GN20/20.2 \citep[$z=4$][]{pop05,dad09b}.  A reasonable inference is
that some $z>4$ DSFGs have warmer dust temperatures, and hence are not
selected by our method.

\citet{Casey:12} performed an optical redshift survey of {\it
  Herschel}-selected sources, some of which were 500\,\mum-risers.  In
order to be included in this sample, a 24\,\mum\ or radio counterpart
was required, which is expected to bias against high-$z$ sources.
Furthermore, the spectroscopic success rate also has a redshift
dependence, and it is possible that the counterpart-identification
process may have a higher mis-identification rate for high-$z$
sources.  Together, these effects make it somewhat difficult to
compare this sample to ours. Using the flux densities derived as part
of the counterpart identification process, there are two sources that
would satisfy our selection criteria.  One of those is clearly an
error -- there is no detected source in any of the SPIRE bands.
Analysis of the maps at the location of the other source shows that it
would barely fail our selection criteria in \fplw\ and $D$, but that
the source appears to be red.  It has a optical redshift of 0.47.  It
is possible that this is a lensed source, and that the redshift is that of
the foreground lens; sub-mm/mm interferometry would be required to
be sure.

\citet{vieira:13, weiss:13} present the results of a redshift survey
of SPT-selected DSFGs from 1300 deg$^2$ of observations.  The
selection requirements are $S_{1.4\,\mathrm{mm}} > 20$\,mJy and that
the 1.4\,mm vs.\ 2\,mm flux density ratio is dust-like rather than
synchrotron-like, with additional screening against $z<0.03$
candidates and radio loud sources.  They obtained secure, multi-line
spectroscopic redshifts, primarily from the $^{12}$CO ladder, for 19
sources, as well as single- or low signal-to-noise multi-line,
redshifts for an additional six objects.  Furthermore, they obtained
{\it Herschel}/SPIRE observations of this sample.  By selection, these
sources are significantly brighter than the sample presented here, and
therefore, as expected from models \citep[e.g.,][]{negrello07}, follow
up reveals that all of these sources are strongly lensed.  This is a
well defined sample with SPIRE coverage, and hence is useful to
compare with our study.

While the authors argue that the redshifts of several of the SPT
sources with single line detections can be inferred from the shape of
the thermal SED (by assuming ``typical'' dust temperatures) or the
lack of additional lines, the possibility of weak CO lines or unusual
dust temperatures -- examples of which are known from the literature
\citep[e.g.,][]{Conley:11, Casey:12} -- remains.  Therefore, here we
consider only SPT sources with secure, multi-line spectroscopic
redshifts.  With this requirement, the band coverage of the redshift
search (primarily from 84 to 115 GHz) imposes a somewhat complicated
redshift selection effect which favors $z>3$ -- that is, for sources
at $z<3$ only single lines could be detected in most cases at the
depth of their observations, resulting in ambiguous redshifts.  For
some of the SPT sources this is alleviated by the presence of
additional follow-up observations at other frequencies.  Furthermore,
the addition of the 1.4\,mm flux selection clearly introduces a
relative selection bias between the SPT sources and the sample
presented here, even for SPT sources that are also 500\,\mum-risers --
although the amount can not be estimated without adding strong
assumptions about the underlying population.  Detailed lens models are
available for four of the SPT sources, but if we assume lens
magnification factors of 5--25, as suggested by those four sources,
then, on average, the SPT population is intrinsically less luminous
than our sample, assuming that our sources are not highly lensed on
average.  Our method is substantially more efficient at selecting
high-$z$ sources (in terms of the overall fraction of sources at
high-$z$ as well as the source density), but, because they are
strongly lensed, the SPT sources are much easier to study in detail.

Ignoring these complications, seven of the 10 SPT sources with $z>4$
are 500\,\mum-risers, while the other three are brightest at
350\,\mum.  One of these is clearly not red in the SPIRE bands, while
the other two have $\fplw \simeq \fpmw$ to within 1$\sigma$.  In
addition, there are three SPT 500\,\mum-risers with $3 < z < 4$ and an
additional four 500\,\mum-risers with single line redshifts which we
exclude from this discussion.  This suggests that red SPIRE colors are
a relatively efficient and complete method of finding 1.4\,mm-selected
sources at $z>4$.


\section{Interpretation}
\label{sec:interpretation}
\subsection{Comparison of Observed Number Density with Galaxy Evolution Models}
\label{subsec:modelcompare}
\nobreak 
To compare the number of observed bright and red SPIRE sources to the
number predicted by pre-existing models, we consulted models by
\citet{bethermin11, bethermin:12, fernandez08, franceschini10,
  leborgne09, valiante09, xu01} using mock catalogs.  The comparisons
are summarized in Table~\ref{tbl:models} and discussed in detail in
this section.  Generally, these models are not physically motivated,
and are developed based on the properties of lower-$z$ DSFGs.  The
goals of this comparison are to test if the properties of the DSFG
population can be simply extrapolated from $z \sim 2$, or if
additional evolution is required, and how well these models can be
used to plan for future surveys of high-$z$ DSFGs.

For the \citet{fernandez08} simulations, which are based on
\citet{Lagache:04}, we used a simulated data set covering 5.2\,deg$^2$
and with a bias parameter 1.5.  For the \citet{leborgne09} model, we
used a mock catalog covering 10\,deg$^2$ and containing 2 million
galaxies.  In both cases we found zero sources satisfying our
selection criteria.  We can therefore conclude that the predictions of
these models for the number of red sources are overwhelmingly ruled
out by our observations.

The other models predict some red sources, but in general they do not
match the properties of our observed population.  A 93\,deg$^2$
simulation of the B11 model, containing 20 million galaxies, predicts
1.1\,deg$^{-2}$ red sources satisfying our selection criteria, but only
0.05 deg$^{-2}$ lie at $z>3$ and 0 at $z > 4$, corresponding to a 95\%
upper limit of $< 0.03$ deg$^{-2}$.  Hence, this model over-predicts
the number of low-$z$ sources, and significantly under-predicts the
number of high-$z$ ones.  The \citet{valiante09} model, which is based on
observed galaxies with luminosities up to 10$^{12.8}$ $\mathrm{L}_{\odot}$ and
incorporates a correlation of luminosity and mean dust temperature,
predicts 15 sources satisfying our selection criteria in a simulated
area of 10\,deg$^2$ and containing 8.2 million galaxies.
However, zero of these galaxies are at $z > 4$, and only three are at $z >
3$.  Furthermore, they have significantly colder dust temperatures
($\sim 30$\,K) than our sources with known redshifts.

The \cite{franceschini10} model over-predicts the number of sources
that meet our observational selection criterion, predicting
10\,deg$^{-2}$.  However, their predictions for the number of sources
at high-$z$ are closer to our limits: 1.2\,deg$^{-2}$ at $z>4$ and 
3.4\,deg$^{-2}$ at $z>3$.  Thus, their prediction for $z>4$ is only
lower than the observed number by a factor of 2 (1.5$\sigma$),
although they also predict a large number of red sources at lower $z$
that are not consistent with observations.

For the \citet{xu01} models, we used mock catalogs covering
10\,deg$^2$ which have been updated to better match the Spitzer number
counts. This substantially over-predicts the number density of red sources at
12\,deg$^{-2}$.  However, the predicted number of $z>4$ sources
is much smaller, 0.08\,deg$^{-2}$, which is strongly ruled out.
Again, this model also predicts a very large number of red sources at
lower $z$ which are not seen in the data.  Inspection of these sources
show that most are classified as having ``AGN'' rather than
``starburst'' templates.  Closer inspection of the templates that meet
our source selection criteria indicate that the models may have poor
applicability for these types of sources.  For example, the template
within the model catalogs which most often passes the selection
criteria is based on the $10^{11.8}\,\mathrm{L}_{\odot}$ galaxy Mrk~0309,
which is assumed to have a dust temperature of $\sim$22\,K (for $\beta$
= 1.5), despite being scaled up by a factor of 5 in luminosity for the
catalogs, and despite having no published measurements for the
template fit at $\lambda > 100$\,\mum .  Galaxies with luminosity of
10$^{12.5}\,\mathrm{L}_{\odot}$ and average dust temperature 22\,K may exist,
but the Mrk~0309-based template is not yet a convincing example of one.

\begin{figure}
\epsscale{1.1}
\plotone{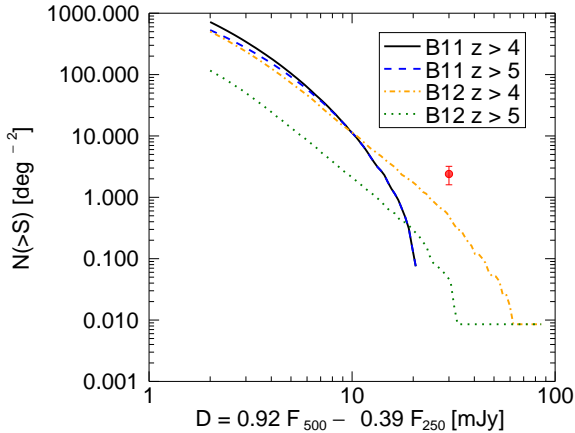}
\caption{The cumulative number counts (red DSFGs brighter than a
  specified flux density in the difference map, $D$) for the B11 and
  B12 models, based on the simulations discussed in the text.  Shown,
  for comparison, is our measurement of the number density of $z>4$
  red DSFGs from \S\ref{subsec:sourcedens}. \label{fig:modeln} }
\epsscale{1}
\end{figure}

The B12 models fare considerably better.  We generated catalogs
representing 200\,deg$^2$, and found a predicted red source density of
0.58\,deg$^{-2}$.  Furthermore, all of the red sources predicted by
this model are at $z>3$, with 0.49\,deg$^{-2}$ at $z>4$.  Folding in
the efficiency distribution, and including the predicted contamination
rate by ``orange'' sources, the B12 model predicts that we should
select 14.7 sources over our fields, compared with the 38 actually
found.  This corresponds to a $P$ value of $3 \times 10^{-7}$, so
formally this model is still excluded at very high significance.
However, it is clear that it is much closer to our observations than
the other models, and it is worth attempting to understand why this is
the case.  There are two primary reasons.  First, the B12 predicts
more high-$z$, luminous DSFGs than many of the other models --
although not, for example, the B11 model. Second, the SED templates
for luminous DSFGs are redder, on average, than for most of the
other models -- this is the most important difference in comparison
with the B11 model.  In addition, the B12 model implements some
scatter in the templates for a fixed $z$ and luminosity, which most of
the other models do not.  However, this seems to have a relatively
modest effect -- doubling the scatter increases the predicted source
density by about 25\%, but entirely by adding bright $z < 3$ red
sources which, so far, have not been observed.  Furthermore, it
disturbs the agreement of this model with the observed far-IR/sub-mm
monochromatic differential number counts.  The cumulative number counts
for the B11 and B12 models, compared with our observations, are
shown in Figure~\ref{fig:modeln}.

One possibility for improving the agreement of these models with
observations is that there may be significantly more strong
lensing at high redshifts than expected.  Here we explore how much
{\it more} lensing would be required to match our observations with
current population models using the B11 model.  The B11 model already
contains lensing, but we can consider the effects of modifying this
prescription.  If we substitute the lensing model of
\citet{wardlow:13}, the number of red sources is significantly
increased to 1.4 deg$^{-2}$.  However, most of these are at $z<2$; the
number densities at $z>3$ and $z>4$ are increased to 0.3 and 0.2
deg$^{-2}$, respectively.  This is a step in the right direction, but
is clearly still inconsistent with our observations.  Furthermore, the observed
flux density distribution clearly does not match our population -- the
number of $\fplw > 50$\,mJy sources at $z>3$ is only 0.05\,deg$^{-2}$.
The five sources with redshifts that we have, even ignoring efficiency
corrections, already rule this out at $>99.8\%$.  

Next we consider modifying the \citet{wardlow:13} lensing model.  If
we scale up all lensing probabilities above $\mu=1.5$ by a constant
factor, we find that we must increase them by 20$\times$ in order to
match our source density at $z>3$, which is extremely unlikely.
Rather than increasing the frequency of large magnifications, which
does not seem to work, a different way to look at the problem is to
impose a minimum magnification floor at $z>3$ on the model.  We can
not do so for all sources, because for physical reasons the mean
magnification over all lines of sight must be one to any redshift.
However, because large magnifications are rare in any such model, a
reasonable approach is to adjust the \citet{wardlow:13} model by
setting $P\left(\mu\right) = 0.5$ between $\mu = 1$ and
$\mu=\mu_{\mathrm{low}}$ (with the probability distribution above 
$\mu_{\mathrm{low}}$ unaltered), and explore how large $\mu_{\mathrm{low}}$ must be
to reconcile models with observations.  Carrying out this procedure,
we find that, for the B11 model, we require $\mu_{\mathrm{low}} =
2.1$.  In other words, {\it half} of all sight-lines to $z>3$ must be
magnified by a factor of two or more.  This is not realistic.
Applying the same analysis to the B12 model does not change much --
to match observations either the high-magnification tail must be 
increased by 15$\times$, or half of all sight-lines must be magnified
by more than $\mu = 1.8$.  

As discussed in \S\ref{subsec:sourcedens}, our measured number density
is effectively at the resolution of SPIRE ($>18\arcsec$).  While our
counts are corrected for blends with non-red objects, if our sources
have a high multiplicity (i.e., all of our catalog objects are in fact
composed of two or more red sources), then this will affect the
\mbox{(dis-)agreement} with models.  The best way to investigate this
topic is to obtain higher-resolution (interferometric) sub-mm/mm
observations.  However, because the above phenomenological models are
trained to match observed number counts that have the same resolution
issues, in order for multiplicity to bring their predictions in
agreement with our observations it must be a strong function of
redshift such that it affects our sources much more than lower-$z$
DSFGs.  This does not seem particularly likely, but it can not be
ruled out at this point.  For example, if luminous DSFGs at all
redshifts are actually comprised of two sources too close to each
other to be resolved, then, in order to maintain agreement with the
overall observed number counts of all sources, the number of predicted
red (but now undetectably blended) sources at SPIRE resolutions would
not change.

Overall, pre-existing models significantly under-predict the number of
high-$z$, red sources we detect.  The one exception is the model of
\cite{franceschini10}, which is a reasonable match for the number of
high-$z$ galaxies detected by our technique, but also predicts a very
numerous population of lower-$z$ red sources highly inconsistent with
{\it Herschel} observations.  Of the models that do not share this
issue, the B12 model performs the best, although it is still ruled out
at high significance.  In any case, this comparison suggests that
there must be additional evolution in the DSFG population at $z>3$
beyond that predicted by simple extrapolations from lower redshift.
It seems certain that gravitational lensing will play a role in
reconciling population models with observations, but for the current
generation of such models the required amount of lensing is not
reasonable.  


\begin{table*}[t!]
 \centering
 \caption{Predicted number of red SPIRE sources}
 \begin{tabular}{l|ccc|l}
 \hline
       & \multicolumn{3}{c}{Predicted Density [deg$^{-2}$]} & \\
 Model & Red & Red $z>3$ & Red $z>4$ & Reference \\
 \hline\hline
B{\'e}thermin 11 & 1.1   & 0.05     & $<0.03$     & \citet{bethermin11}; B11 \\
B{\'e}thermin 12 & 0.58  & 0.58     & 0.49        & \citet{bethermin:12}; B12\\
F.-Conde      & $<0.6$   & $<0.6$   & $<0.6$      & \citet{fernandez08} \\
LeBorgne      & $<0.3$   & $<0.3$   & $<0.3$      & \citet{leborgne09} \\
Valiante      & 1.5 & 0.2           & $<0.3$      & \citet{valiante09} \\
Franceschini  & 10  & 3.4           & 1.2         & \citet{franceschini10} \\
Xu            & 12  & 0.34          & 0.08        & \citet{xu01} \\
\hline
Observed      & $3.27^{+0.67}_{-0.84}$ & $>1.68$ (95\%) & $2.37^{+0.75}_{-0.79}$ & 
 This work \\
\end{tabular}
\tablecomments{Comparison of the prediction of various pre-existing
  models with the observed number of red sources satisfying our
  detection criteria ($D \ge 23.9\,\mathrm{mJy}, \fplw \ge 30\,\mathrm{mJy},
  \fplw \ge \fpmw \ge \fpsw$).  
  The upper limits, which are provided when no
  such sources were generated in our simulations, are the $95\%$
  one-sided frequentist limits, and are set by the sky area simulated
  for each model.  The number of red sources at $z>3$ and $z>4$ are
  based on combining the measured sky density with the observation
  that 4/5 of our sources with redshifts are at $z>4$, and 5/5 at
  $z>3$. \label{tbl:models}}
\end{table*}


\subsection{Overlap of Our Red SPIRE Sample with Millimeter-Selected Galaxies}
\label{subsec:mmoverlap}
\nobreak
Given our selection criterion of monotonically rising spectra with
increasing wavelength in the SPIRE bands, we would expect the selected
objects to be strong emitters at $\lambda \approx 1\,\mathrm{mm}$.
This is now demonstrated for a subset of the sample, as discussed in
\S\ref{sec-followup}.  In this section, we examine the reverse
implication using mm-wave observations that cover a small fraction of
the HerMES fields: do mm-selected sources have properties similar to
the red SPIRE sources in this paper?

With two exceptions\footnote{The South Pole Telescope
  \citep{mocanu:13} and Atacama Cosmology Telescope \citep{marsden:13}
  surveys; due to their large areas, however, these are effectively
  much shallower than the HerMES survey.}, existing $\lambda \approx
1$\,mm ``blind'' surveys sensitive enough to detect significant
numbers of high-redshift galaxies cover areas less than 1 deg$^2$, and
measure a population with $\sim 100\times$ the surface density of our
red SPIRE galaxies.  Therefore, by virtue of observational selection,
we can not have good correspondence with these surveys.  It is
reasonable to expect that the degree-sized mm surveys will uncover
objects that are typically fainter at far-IR through millimeter
wavelengths and likely have lower luminosity.  The comparison with
lensed, mm-selected SPT sources (which also have SPIRE observations)
was already presented in \S\ref{sec:otherred}.

Within the SPIRE SDP fields discussed in this paper, the AzTEC 1.1\,mm
survey of Lockman Hole East \citep{austermann10} covers the largest
area with $\sim$1\,mJy depth.  The AzTEC survey produced a primary
catalog of 43 sources with $4\sigma$ statistical significance over an
area of 0.3\,deg$^2$.  None of these correspond to red SPIRE sources
cataloged in Table~\ref{tbl-src}.  However, all 17 of the AzTEC
sources with greater than $5\sigma$ statistical significance are
associated with clear SPIRE sources and positive peaks in the
difference image $D$.  These sources range in 1.1\,mm brightness from
3.6 to 6.6\,mJy (deboosted), with a mean of 4.6\,mJy.  We stacked the
SPIRE 50\arcsec-resolution images at the positions of the 17 AzTEC
sources.  The stacked mean values of $\fpsw, \fpmw, \fplw$ and $D$ are
30.1, 34.1, 25.7, and 11.4 mJy, respectively.  Collectively, then,
these mm-selected sources peak at shorter far-IR wavelengths than the
red sample in this paper, and they are a factor of $\sim$1.3 fainter
in the far-IR.

Only one source -- \#13 -- among the 17 AzTEC $5\sigma$ sources has a
SPIRE counterpart which meets our redness criterion, but not our
brightness criterion.  In matched 50\arcsec\ beams, it has flux
densities $S_{250\,\mum}, S_{350\,\mum}, S_{500\,\mum}$ of 12.4,
25.8, 29.0\,mJy, respectively.  A tentative photometric redshift of 2.1
-- ironically below the median for the sample -- was assigned to
source 13 by the AzTEC team \citep{michalowski:12} based on a radio and 
24\,\mum\ counterpart.  However, the radio and 24\,\mum\ source
appear to be offset by similar distance and direction from both the
AzTEC and SPIRE source, so may not be physically associated.

\subsection{Contribution of Red Sources to Far-IR Background}
\nobreak 
Using our completeness and purity simulations, it is simple
to estimate the contribution of sources selected by our method to the
cosmic far-infrared background (CIB) at 500\,\mum\ by weighting the
number density calculation by \fplw .  Carrying out this procedure, we
find for the FLS field a value of $0.66 \pm
0.20\,\mathrm{kJy}\,\mathrm{sr}^{-1}$, and for Lockman-SWIRE $0.38 \pm
0.10\,\mathrm{kJy}\,\mathrm{sr}^{-1}$, where the uncertainty in the
observed fluxes and Poisson fluctuations (which account for $\sim 2/3$
of the error budget) are included, but not the uncertainty in the
SPIRE calibration.  These differ by $1.3\,\sigma$, and hence are
consistent.  Compared with the FIRAS measurement of $0.39 \pm 0.10\,
\mathrm{MJy}\,\mathrm{sr}^{-1}$ \citep{fix98}, the red sources with
$\fplw > 30$ mJy account for only 0.1--0.2\% of the background. Note
that these only represent the intrinsically most luminous high-$z$
DSFGs, and non-red sources at these redshifts are not included, but,
assuming that the luminosity function for DSFGs has a similar shape at
$z>3$ as it does at $z \sim 2$, 500\,\mum-risers probably do not account
for more than a few percent of the CIB at 500\,\mum.

\subsection{Photometric Redshifts}
\label{subsec:pz}
\nobreak 
Up until this point in the paper, we have tried to avoid any analysis
which requires photometric redshifts, but they are required in order
to estimate the contribution of our sources to the star formation
history of the Universe. We have firm spectroscopic redshifts for only
five of the galaxies in our sample.  Without precise interferometric
positions, we are limited to the properties of the thermal dust SEDs
to derive photo-$z$s.  Because modified blackbody models are perfectly
degenerate in $T_\mathrm{d}/\left(1+z\right)$ and $\lambda_0
\left(1+z\right)$, deriving photo-$z$s from these data essentially
requires assuming a prior on the rest frame values of these
parameters.  Here we attempt to derive a crude estimate of the
redshift distribution of our population using such a prior.  Note that
here we are not making use of the priors discussed in
{\S\ref{sec:basic}} in any way.

Instead of working with the dust temperature $T_\mathrm{d}$, which is
rather indirectly constrained by the data, we use the wavelength at which
$S_\nu$ peaks, $\lambda_{\mathrm{max}}$, for each source.  We construct
a prior for the rest-frame $\lambda_{\mathrm{max}}$ by following the
approach of \citet{Greve:12}: collect a comparison sample of DSFGs
(with precise spectroscopic redshifts and well constrained thermal
SEDs), and analyze them using the same modified blackbody model.  If
the (observer frame) peak wavelength of each red source is
$\lambda^S_{\mathrm{max}}$, and the (rest frame) peak wavelength of
each comparison source is $\lambda^C_{\mathrm{max}}$, then the
photometric redshift estimate $z_{\mathrm{p}}$ for that red source
from that comparison is $1 + z_{\mathrm{p}} = \lambda^S_{\mathrm{max}}
/ \lambda^C_{\mathrm{max}}$.  Because our SED analysis is based on
MCMC methods, we can also easily include measurement uncertainties in
each $\lambda_{\mathrm{max}}^{ \{ C,S \} }$.  

The results depend sensitively on the comparison sample.  Ideally,
this sample would be selected in exactly the same manner (adjusting
for redshift) as our red sources -- but no such sample is currently
available.  We considered several possible compilations, evaluating
them based on how well they predict the redshifts of a test sample
consisting of our five red sources with spectroscopic redshifts as
well as the red sources from \citet{combes12} and \citet{cox11}, both
of which were discovered in {\it Herschel}/SPIRE observations and
which would clearly be selected using our method.  We do not include
the SPT red sources in our test sample because it is not clear how the
1.4\,mm selection will bias the selection relative to ours.
\citet{Magnelli:12} present a detailed study of ``classic SMGs'' at
$z\sim2.4$ selected based on longer wavelength
($850\,\mum$--$1.2\,\mathrm{mm}$) observations, and with redshifts
generally determined on the basis of radio cross-identifications,
using the same SED model. The longer wavelength selection is expected
to bias this sample relative to ours towards colder sources, and hence
towards lower redshifts for fixed SPIRE colors.  Indeed, analysis of
this sample gives $z_{\mathrm{p}} \sim 2$ for our seven test sources
($z_{\mathrm{p}} \sim 3$ for FLS 3), which can be ruled out at very
high significance, as 6/7 are at $z>4$.  Therefore, as a comparison
sample this does a very poor job predicting the redshifts of our
sources -- so poorly, in fact, that it is not possible to re-weight it
to provide a good estimate.  The same is true, but to a lesser extent,
for the SPT sample of \citet{vieira:13}: the sources are colder, and
hence under-predict the redshifts of our test sample.

The \citet{Casey:12} study discussed in \S\ref{sec:otherred} presents
redshifts for a large sample of {\it Herschel}-selected DSFGs, with
the additional requirement that they be detected in {\it Spitzer}
24\,\mum\ or radio observations.  While that paper uses a somewhat
different SED model, they also present $\lambda_{\mathrm{max}}$
measurements.  At the typical $L_{\mathrm{IR}}$ of our sources ($\sim
10^{13}\,\mathrm{L}_{\odot}$), this predicts
$\left<z_{\mathrm{p}}\right> \sim 9$, which is again strongly ruled
out.  This may be the result of selection biases in the
$\lambda_{\mathrm{max}}$-$L_{\mathrm{IR}}$ relation of the
\citet{Casey:12} sample.  If we instead select $\fplw > \fpmw > \fpsw$
sources -- now dropping the minimum flux requirements to $\fplw > 10$
mJy in order to increase the comparison sample size (to 19 sources) --
we find $\left<z_{\mathrm{p}}\right> = 6.3$ for the comparison sample
(excluding FLS~3), an improvement, but again strongly ruled out.  These
sources have similar colors as those of our red sample, so this is
somewhat surprising -- the 24\,\mum / radio and optical spectroscopic
success requirements are expected to bias this sample towards lower
$z$, increasing the rest-frame $\lambda_{\mathrm{max}}$ values, which
would lead to an under-prediction of $z_{\mathrm{p}}$ for our test
sources rather than the observed over-prediction.  In any case, this
does not seem to be a good comparison sample for our purposes.

Finally, we consider the compilation of well-observed {\it
  Herschel}-selected lensed sources of \citet{Greve:12}, also at
$z\sim2$ (which these authors used to make photo-$z$ estimates for
1.4\,mm-selected SPT sources).  They fit the photometry of the lensed
sources using a similar SED model, but holding $\beta$ and $\lambda_0
\left(1+z\right)$ fixed.  Because these sources are generally quite
well observed, we re-fit this sample using our model but without
fixing these parameters (or using priors), excluding the few sources
where the data quality was insufficient to constrain such a fit.
Comparison with our test sample gives $\left<1+z_{\mathrm{p}}\right> /
\left<1+z_{\mathrm{spec}}\right> = 0.93 \pm 0.08$, much better than
the other comparison samples.  These highly magnified sources almost
certainly probe a different (fainter) luminosity range than our red
sources.  This may result in some bias in the photo-$z$s; for example,
if dust temperature increases with luminosity, we would expect the
lower-$z$ lensed galaxies to peak at longer rest-frame wavelengths
than our red sources, and hence under-predict the redshifts.  However,
we do not apply any correction, since the derived photo-$z$ values are
consistent overall with the spectroscopic redshifts of our test
sample.  We further supplement this collection by including the fits
to FLS~1, FLS~3, FLS~5, and LSW~20 as in \S\ref{subsec:detailed-fits},
resulting in a final comparison sample of 19 DSFGs.

\begin{figure}
\epsscale{1.1}
\plotone{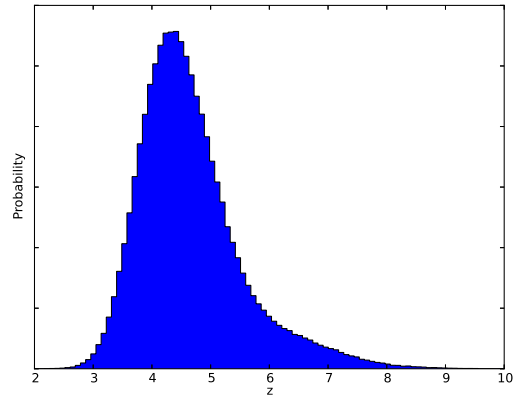}
\caption{The photo-$z$ distribution of our sample using the SED
prior discussed in the text. \label{fig:pz}}
\epsscale{1}
\end{figure}

Applying this effective prior to the $\lambda_{\mathrm{max}}$ values derived in
\S\ref{sec:basic} results in the population distribution given in
Figure~\ref{fig:pz}.  This gives $\left<z_{\mathrm{p}}\right> = 4.7$ with a one
sigma range of $\pm 0.9$, and with 20\% of the population at $z < 4$
-- consistent with the existence of LSW 20 at $z=3.4$ in our test
sample.  The tail to higher redshift is due to the fact that, for the
redder sources in our sample, the upper limit on $\lambda_{\mathrm{max}}$ is
not as well constrained as the lower limit, although the existence of
FLS 3 at $z=6.34$ demonstrates that there are at least some $z>6$ DSFGs.
Adding more longer-$\lambda$ observations would help constrain this
tail.

\subsection{Contribution of these Sources to high-$z$ Star Formation}
\label{subsec:contribsfr}
\nobreak At this time, the paucity of spectroscopic redshifts of our
sources prevents a precise calculation of their contribution to the
star formation history of the Universe.  However, we can extend the
photometric redshift model discussed in \S\ref{subsec:pz} to form a
rough estimate.  Each of the comparison sample sources has a well
determined thermal SED, so in addition to using randomly sampled
$\lambda^C_{\mathrm{max}}$ values, we also take the corresponding
values of the SED parameters ($T_{\mathrm{d}}$, $\beta$, etc.),
redshifting that SED to the corresponding $z_{\mathrm{p}}$ of the
catalog source, and scaling it to match the observed 500\,\mum\ flux
density.  For this calculation we must also correct for the selection
efficiency and purity, as discussed in \S\ref{sec:numdens}, so for
each $z_{\mathrm{p}}$, $L_{\mathrm{IR}}$ pair we also associate a
randomly drawn $\epsilon$ and $p$ appropriate to that source.  Doing
this once for each source in our catalog provides a single simulation
of our red source sample.  We then bin the simulated sources by
$z_{\mathrm{p}}$ into two broad bins ($z_{\mathrm{p}}=$4--5 and
$z_{\mathrm{p}}=$5--6), add up the total $L_{\mathrm{IR}}$ in each
bin, counting each catalog source as $p / \epsilon A$ sources per
deg$^2$, where $A$ is the area of the field that that source comes
from.  For those sources in our main survey area with spectroscopic
redshifts (FLS~1, FLS~3, FLS~5, and LSW~20) we instead sample
$L_{\mathrm{IR}}$ directly from their SED fits
(\S\ref{subsec:detailed-fits}), hence excluding both FLS~3 and LSW~20
from our calculation as they lie outside both redshift bins.

\begin{figure}
\epsscale{1.1}
\plotone{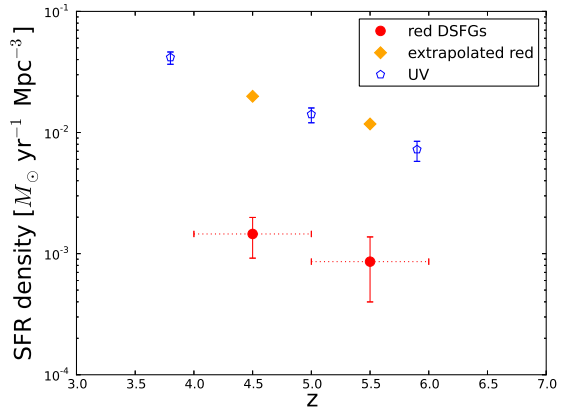}
\caption{Estimated contribution of our 500\,\mum-riser selected DSFGs
  to the star formation rate density at $z=$4--6 (red circles).  No
  correction is made for fainter sources, or for DSFGs in this
  redshift range that are not red in the SPIRE bands.  The horizontal
  bars reflect the bin size, and are not uncertainties.  The orange
  diamonds show the red points corrected for fainter sources using the
  B11 luminosity function.  For comparison, the blue pentagons are the
  extinction corrected values derived from rest-frame UV {\it HST}
  surveys for sources brighter than $0.3 L_{\star}$
  \citep{Bouwens:07}; these include significantly lower luminosity
  sources than the red points. \label{fig:madau}}
\epsscale{1}
\end{figure}

We then divide the total $L_{\mathrm{IR}}$ in each bin by the comoving
volume per deg$^2$, and convert to a star formation rate density
(SFRD) using the \citet{kenn98} conversion.  By repeating this
procedure 5000 times, and folding in Poisson noise due to the limited
sample size, we can estimate the uncertainty in the SFRD from these
sources, deriving values of $\left(1.5 \pm 0.5\right) \times 10^{-3}$
and $\left(8.6 \pm 4.9\right) \times 10^{-4}$ $\mathrm{M}_{\odot}\,
\mathrm{yr}^{-1}\, \mathrm{Mpc}^{-3}$ at $4 \le z < 5$ and $5 \le z <
6$, respectively (Figure~\ref{fig:madau}).  About half of the
uncertainty budget in the lower bin and the majority in the higher bin
arises from Poisson noise.  Note that we make no attempt to correct
for non-red DSFGs at these redshifts, nor for sources fainter than our
detection limit.  These values thus represent only the contribution to the
SFRD from the most luminous, heavily obscured far-IR galaxies.  

This discussion assumes that AGN activity is not a major contributor
to the far-IR luminosity of our sources, as is thought to be the case
for lower-$z$ DSFGs \citep{Alexander:05}.  Unfortunately, given the
high redshift, extreme obscuration, and modest source density of our
sources, obtaining sufficiently deep X-ray data for even a small
fraction of our catalog would be prohibitively expensive.  However, we
note that the closest X-ray source to GOODSN~8 in the 2\,Msec {\it
  Chandra} Deep Field North catalog \citep{Alexander:03} is
18\arcsec\ away, which, given our positional uncertainties, is
unlikely to be related.

If we assume that the {\it shape} of the DSFG luminosity function does
not evolve from $z \sim 2$ to $z \sim 5$ (which is almost certainly
not true in detail, and may not even be a good approximation), then we
can correct for the presence of fainter starbursts.  Further assuming
that our $z>4$ targets have $L_{\mathrm{IR}} >
10^{13}\,\mathrm{L}_{\odot}$, based on the results of
\S\ref{subsec:detailed-fits}, and using the B11 luminosity
function \footnote{See Figure~11 of B11, noting that the only change
  in the shape from $z=3$ to $5$ is an increase in $L_{\star}$ by
  6\%.}, the contribution to the SFRD from red, $z>4$ DSFGs is similar
to the extinction-corrected, UV-inferred SFRD at the same redshifts of
\citet{Bouwens:07}, as shown in Figure~\ref{fig:madau}.  Using the B12
luminosity function increases this by a further factor of three.
Again, we have not made any attempt to include non-red $z>4$
DSFGs. Furthermore, the overall SFRD for these luminosity functions is
dominated by sources with $L_{\mathrm{IR}} \sim
10^{12}\,\mathrm{L}_{\odot}$, which are too obscured to be detected in
the rest-frame UV even in deep {\it HST} observations.  Clearly, the
exact values should not be taken too seriously, but they do suggest
that rest-frame UV based estimates of the star formation history of
the Universe may be missing a significant component of the SFRD, even
after corrections for extinction.  Determining whether or not this is
the case will require further observations -- with the goals of both
increasing the number of red sources with spectroscopic redshifts, and
extending the search to fainter sources.

\section{Conclusions}
\nobreak 
We have presented a method for selecting candidate high-$z$ dusty
star-forming galaxies using {\it Herschel}/SPIRE colors, and provided
a catalog of such sources selected from the first 21.4 deg$^2$ of data
from the HerMES project.
\begin{itemize}
\item The number density of red $\fplw \geq 30, D = 0.92 \fplw - 0.39
  \fpsw \geq 23.9$ mJy sources in confusion limited SPIRE maps is
  $\sim 2$ deg$^{-2}$. After modeling for selection efficiency and
  contamination, this implies an underlying number density of $3.3^{+0.7}_{-0.8}$
  deg$^{-2}$.  The 95\% lower limit is 2.0 deg$^{-2}$.
\item We have obtained redshifts for five targets selected using our
  method.  They range from $z=3.4$ to 6.3, with 4/5 at $z>4$.
\item SED modeling of the sources with redshifts shows that they
  correspond to the luminous tip of the lower-$z$ DSFG luminosity
  function, with typical $L_{\mathrm{IR}} > 10^{13}$ $\mathrm{L}_{\odot}$.
\item Combining the measured redshifts with the source density, we
  estimate the number of red, luminous $z>4$ DSFGs to be $2.4 \pm 0.8$
  deg$^{-2}$; the largest uncertainty is due to the small number of
  sources with secure spectroscopic redshifts.
\item The number of such high-$z$ DSFGs is significantly higher than
  the predictions of existing population models.  While gravitational
  lensing clearly has a role to play in understanding this population,
  the amount required to reproduce observations with current population models
  appears to be unrealistic.
\item Using a photo-$z$ model based on the properties of lower-$z$, lensed,
  {\it Herschel}-selected DSFGs, we estimate $\left<z\right> \sim 4.7$
  for our population.  
\item Extending this model to predict the contribution of red sources
  satisfying our selection criteria to the star formation rate
  density, we find $\sim 10^{-3} \,\mathrm{M}_{\odot}\,
  \mathrm{yr}^{-1} \, \mathrm{Mpc}^{-3}$ at $z \sim 5$.  This is
  significantly lower than the total inferred SFRD at these redshifts
  -- but these represent only the tip of the luminosity function.
\item If the ratio between the most luminous sources and the rest of
  the population is similar at $z>4$ to that for lower-$z$ DSFGs, then
  DSFGs may contribute a similar amount to the overall star formation
  density at these redshifts as the extinction corrected, rest-frame UV 
  measurements -- which do not account for such highly extinguished systems.
\end{itemize}
This sample represents less than 1/5 of the 110 deg$^{2}$ of the main
HerMES survey, which reaches similar depths, and an even smaller
fraction of the 300 deg$^{2}$ of the shallower HeLMS companion survey
\citep{hermes12}.  The number density of sources selected by our
method should be a good description of all but the deepest (e.g.,
GOODS-N) confusion limited SPIRE maps, although we are working to
improve this by using more optimal filtering.  In shallower surveys,
such as HeLMS or H-ATLAS \citep{eales:10}, the number density will be
lower.  For example, to a depth of 50 mJy at 500\,\mum, the number
density (without purity or completeness corrections) is $\sim 0.5$
deg$^{-2}$, about $4\times$ less, based on our catalogs.  Follow-up of
our sources is ongoing at a wide range of facilities.  
Obtaining more redshifts is the highest priority for future studies.

\acknowledgements This material is based upon work supported by the
National Aeronautics and Space Administration under Grant
No.\ 12-ADAP12-0139 issued through the ADAP program.  Part of this
work was performed at the Jet Propulsion Laboratory, California
Institute of Technology, under a contract with NASA.  SPIRE has been
developed by a consortium of institutes led by Cardiff Univ. (UK) and
including Univ. Lethbridge (Canada); NAOC (China); CEA, LAM (France);
IFSI, Univ. Padua (Italy); IAC (Spain); Stockholm Observatory
(Sweden); Imperial College London, RAL, UCL-MSSL, UKATC, Univ. Sussex
(UK); Caltech, JPL, NHSC, Univ. Colorado (USA). This development has
been supported by national funding agencies: CSA (Canada); NAOC
(China); CEA, CNES, CNRS (France); ASI (Italy); MCINN (Spain); SNSB
(Sweden); STFC (UK); and NASA (USA).  The SPIRE data presented in paper
this have been released through the {\it Herschel} Datablase in
Marseille, HeDAM (http://hedam.oamp.fr/HerMES).  Some of this material
is based upon work at the Caltech Submillimeter Observatory, which was
operated by the California Institute of Technology under cooperative
agreement with the National Science Foundation (AST-0838261).  The
Submillimeter Array is a joint project between the Smithsonian
Astrophysical Observatory and the Academia Sinica Institute of
Astronomy and Astrophysics and is funded by the Smithsonian
Institution and the Academia Sinica.  Based in part on observations
carried out with the IRAM Plateau de Bure Interferometer. IRAM is
supported by INSU/CNRS (France), MPG (Germany) and IGN (Spain).
Support for CARMA construction was derived from the Gordon and Betty
Moore Foundation, the Kenneth T. and Eileen L. Norris Foundation, the
James S. McDonnell Foundation, the Associates of the California
Institute of Technology, the University of Chicago, the states of
California, Illinois and Maryland, and the National Science
Foundation. Ongoing CARMA development and operations are supported by
the National Science Foundation under a cooperative agreement and the
CARMA partner universities.  Some of the data presented herein were
obtained at the W.M. Keck Observatory, which is operated as a
scientific partnership among the California Institute of Technology,
the University of California and the National Aeronautics and Space
Administration. The Observatory was made possible by the generous
financial support of the W.M. Keck Foundation.  Based in part on
observations made with the Gran Telescopio Canarias (GTC), installed
in the Spanish Observatorio del Roque de los Muchachos of the
Instituto de Astrofisica de Canarias, in the island of La Palma. This
research made use of Astropy (http://www.astropy.org), a
community-developed core Python package for Astronomy
\citep{astropy:13}.  This research has made use of NASA's Astrophysics
Data System Bibliographic Services.  This research has made use of the
NASA/IPAC Extragalactic Database (NED) which is operated by the Jet
Propulsion Laboratory, California Institute of Technology, under
contract with the National Aeronautics and Space Administration.

Facilities: \facility{Herschel (SPIRE)}, \facility{CSO (Bolocam,
  Z-Spec)}, \facility{CARMA}, \facility{SMA},
\facility{IRAM:Interferometer}, \facility{Keck:I (LRIS)},
\facility{GTC (OSIRIS)}, \facility{e-MERLIN}

\appendix
\section{Modified Blackbody SED fits}
\label{apndx:mbb}
Here we discuss the details of our modified blackbody fits to the
thermal SEDs of our sources.  The basic model is given by
Equation~\ref{eqn:mbb}, and is written in terms of the blackbody
function (through the dust temperature $T_{\mathrm{d}}$), the dust properties
($\beta$ and $\lambda_0$), and the overall normalization ($\Omega$).
Note that using $\Omega$ directly is a poor choice for fitting
purposes because it is extremely degenerate with $T_d$, so instead we
use the observer frame flux density at 500\,\mum\ as the normalization
parameter.  Tests show that this parameterization is much better behaved.
We use flat priors over all model parameters.

We have developed code to fit this model to SED data using an Markov
Chain Monte Carlo (MCMC) method.  An earlier version of this code was
also used by \citet{rie13} to analyze the photometry of FLS 3, and we
have made it freely
available\footnote{https://github.com/aconley/mbb\_emcee.}.  We make
use of the parallelized affine-invariant MCMC {\tt python} module {\tt
  emcee} \citep{emcee:12}, which in practice has better convergence
properties than the more commonly used Metropolis-Hastings method.  We
check convergence of our chains by requiring that the autocorrelation
length for each model parameter is several times less than the number
of steps taken during an initial burn-in stage (which are then
discarded).

Compared with the version of this code used by \citet{rie13}, 
several improvements have been made. First, we have added the
ability to impose upper and lower limits as well as Gaussian priors on
the model parameters.  These additions are useful for analyzing poorly
constrained data, such as when we consider fits to only the SPIRE
photometry in \S\ref{sec:basic}.  Second, instead of treating the
instrument response as a delta function in wavelength, we now fully
model the wavelength response of each instrument.  We have applied
this improved code to FLS~3, and find that the resulting changes in
the model parameters are well within the quoted uncertainties.
\vskip 0.1cm



\end{document}